\documentclass[fleqn,usenatbib]{mnras}

\usepackage[T1]{fontenc}
\usepackage{lmodern}
\usepackage{hyperref}

\DeclareRobustCommand{\VAN}[3]{#2}
\let\VANthebibliography\thebibliography
\def\thebibliography{\DeclareRobustCommand{\VAN}[3]{##3}\VANthebibliography}

\usepackage{graphicx}	
\usepackage{makecell}
\usepackage{gensymb}
\usepackage{amsmath,amssymb}	
\usepackage[figuresright]{rotating}
\makeatletter

\let\c@sidewaystable\c@table
\let\c@sidewaysfigure\c@figure
\makeatother

\usepackage{orcidlink}

\newcommand{\sigvod}{\sigma_{v,1\mathrm{D}}}
\newcommand{\sigvtd}{\sigma_{v,3\mathrm{D}}}
\newcommand{\mach}{\mathcal{M}}
\newcommand{\macha}{\mach_\mathrm{A}}
\newcommand{\avir}{\alpha_\mathrm{vir}}
\newcommand{\sigmarho}{\sigma_{\rho/\langle\rho\rangle}}
\newcommand{\pc}{\mathrm{pc}}
\newcommand{\D}{\mathrm{D}}
\newcommand{\CO}{\mbox{CO\,($J$=2--1)} }
\newcommand{\cs}{c_\mathrm{s}}
\newcommand{\SigmaN}{\Sigma/\langle\Sigma\rangle}
\newcommand{\tpeak}{T_\mathrm{peak}}
\newcommand{\SFRff}{\mathrm{SFR_{ff}}}
\newcommand{\tff}{t_\mathrm{ff}}

\definecolor{cornflowerblue}{rgb}{0.39, 0.58, 0.93}

\NewDocumentEnvironment{responselong}{}  
{\color{cornflowerblue}} 

\title[Turbulence in NGC7793 and NGC1313]{The turbulence driving mode in NGC7793 and NGC1313}

\author[L.J.~Miller et al.]{
Lewis J.~Miller,$^{{\orcidlink{0009-0008-8732-8506}}\,1}$ \thanks{E-mail: lewis.miller@anu.edu.au}
Kathryn Grasha, $^{\orcidlink{0000-0002-3247-5321}\,1}$
and Christoph Federrath$^{\orcidlink{0000-0002-0706-2306}\,1}$
\\
$^{1}$Research School of Astronomy and Astrophysics, Australian National University, Canberra, ACT 2611, Australia
}

\date{Accepted XXX. Received YYY; in original form ZZZ}

\pubyear{\the\year{}}

\begin{document}

\label{firstpage}
\pagerange{\pageref{firstpage}--\pageref{lastpage}}
\maketitle


\begin{abstract}
We present spatially resolved measurements of turbulence driving modes across entire extragalactic discs of NGC7793 and NGC1313, using Atacama Large Millimetre/submillimetre Array (ALMA) \CO observations at $13\,\pc$ resolution. By applying a kernel-based analysis of density and velocity fluctuations, we map the turbulence driving parameter, $b$, which characterises the balance between solenoidal ($b\sim0.3$) and compressive ($b\sim1$) turbulent driving regimes. $b$ is quantified as the ratio of the turbulent density fluctuations relative to the turbulent sonic Mach number, $\mach$. Both galaxies show predominantly solenoidal driving on average for the regions where we find valid results ($b\geq0.33(\pm0.05)^{+0.14}_{-0.10}$ in NGC7793; $b\geq0.24(\pm0.03)^{+0.10}_{-0.07}$ in NGC1313), noting that this is without including the influences of magnetic fields, making these measurements lower limits. We find substantial spatial variation of $b$, including localised regions of strongly compressive driving. NGC1313 exhibits higher turbulent Mach numbers and density dispersions than NGC7793, consistent with the disturbed morphology and recent satellite interaction in NGC1313. The turbulence in both NGC7793 and NGC1313 is supersonic ($3\lesssim\mach\lesssim20$), and NGC1313 shows a radially decreasing trend of $\mach$ with galactocentric radius. Radial trends indicate more solenoidal driving in the galaxy centres, potentially reflecting enhanced shear, and increasingly compressive modes in the outskirts. These results demonstrate that turbulence driving varies systematically with galactic environment and cannot be assumed uniform across discs. Our study applies a previously established method to larger scales and new data, linking local turbulence physics to global star formation regulation in galaxies, providing a new avenue for testing theoretical models with future integral field units (IFU) and ALMA surveys.

\end{abstract}

\begin{keywords}
turbulence -- galaxies: star formation -- galaxies: ISM --  ISM: kinematics and dynamics -- stars: formation
\end{keywords}



\section{Introduction}

Stars form from condensing clouds of cold molecular gas within the interstellar medium (ISM). The star formation process is highly complex, governed by multiple physical factors. Since the 1970s, studies have revealed that the gas depletion time in our Galaxy is significantly larger than the free-fall time of any star-forming gas \citep{Zuckerman_1974, Williams_1997}. This discrepancy has since been observed in local gas clouds (within the Milky Way), disc galaxies, and starburst galaxies across various redshift, where gas depletion times are found to be of the order of 100~free-fall times \citep{Krumholz_2007, Krumholz_2012, Federrath_2013}. Such findings indicate that gravity alone cannot account for the observed star formation rates (SFRs), necessitating the consideration of additional regulatory mechanisms. 

Early theories proposed that magnetic fields counteract gravitational collapse of molecular gas clouds, thereby extending gas depletion times \citep{Shu_1987}. However, more recent studies emphasise the role of local ISM properties in determining SFRs \citep{Krumholz_2012}. Local factors, such as gravity, turbulence, magnetic fields, stellar feedback, thermodynamics, and molecular gas chemistry, primarily regulate star formation. These local conditions, in turn, are influenced by large-scale galactic factors, including the stellar potential, rotational dynamics, and accretion. Since the 1990s, advancements in computational capabilities have enabled large-scale, three-dimensional magnetohydrodynamic (MHD) simulations. The focus of research has shifted from magnetic fields to turbulence as a dominant mechanism regulator of star formation and emphasising that while magnetic fields play a role, turbulence significantly affects the dynamics of the ISM and star formation by both opposing and enhancing gravitational collapse \citep{Elmegreen_2004, Mac_Low_2004, McKee_2007, Federrath_2012, Federrath_2015, Herron_2017, Kortgen_2017, Menon_2020_expanding, Kortgen_2020, Dhawalikar_2022, Mathew_2025}. 

These and other studies \citep[e.g.,][]{RenaudEtAl2012, Federrath_2013, SalimFederrathKewley2015, Hennebelle_2024} have shown that four key factors govern the normalised SFR per freefall time ($\SFRff$); the virial parameter ($\avir$), the sonic Mach number ($\mach$), the plasma beta ($\beta = 2 \macha^2 / \mach^2$, where $\macha$ is the Alfvén Mach number) and the turbulence driving parameter ($b$) \citep{Federrath_2012}. The central reason for the SFR dependence on these parameters is that the dense-gas fraction depends on the standard deviation of the mean-normalised volumetric turbulent density fluctuations, $\sigmarho$, which in turn is largely determined by three of these parameters: $\mach$, $\beta$, and $b$, such that
\begin{equation} \label{eqn:b_def_magnetic}
    \sigmarho^2 = b^2\mach^2 \frac{\beta}{\beta+1}.
\end{equation}
If the magnetic field is assumed to be zero (i.e., $\beta \to \infty$), then the factor $\beta/(\beta+1)\to1$. We discuss the influence of this assumption in Sec.~\ref{sec:caveats}. Effectively, the assumption means that $b$ represents a lower limit for a given measurement of $\sigmarho$ and $\mach$. In this limit, rearranging for the turbulence driving parameter we find
\begin{equation} \label{eqn:b_def}
    b = \frac{\sigmarho}{\mach}.
\end{equation}

The turbulence driving parameter characterises the driving mode of turbulence in the ISM and significantly influences star formation. Despite its importance, measuring $b$ is nontrivial, and it has not been directly observed outside the Milky Way and its satellites. The parameter $b$ defines the mode of turbulence driving the gas, ranging from purely solenoidal (divergence-free; $b=1/3$) to purely compressive (curl-free; $b=1$), supported by both theoretical modelling \citep{Molina_2012} and hydrodynamic simulations \citep{Federrath_2010}. Consequently, measuring $b$ provides insight into the dominant turbulence driving mode, a key physical process shaping the ISM and star formation dynamics.

Several numerical studies have explored the turbulence driving parameter in different, idealised and controlled environments. Recent simulation results \citep[e.g.,][]{Herron_2017, Kortgen_2017, Menon_2020_expanding, Kortgen_2020, Dhawalikar_2022} underscore how the driving mode of turbulence shapes the structure and dynamics of the ISM, as well as regulating the star formation. These MHD simulations explore various galactic environments, providing valuable insights into the role of turbulence in galactic kinematics. 

Observationally, the turbulence driving parameter has been measured in various Milky Way environments, including the Taurus molecular cloud \citep{Brunt_2010_taurus}, a non-star-forming cloud along the line of sight towards W49A \citep{Ginsburg_2013}, the central molecular zone (CMZ) cloud known as the `Brick' \citep{Federrath_2016_Brick}, clouds in the solar neighbourhood \citep{Kainulainen_2017}, pillars in the Carina Nebulae \citep{Menon_2020}, and neutral hydrogen gas clouds entrained in the hot nuclear wind of the Milky Way \citep{Gerrard_2024}. These investigations provide valuable insights into turbulence under varying physical conditions across the Galaxy. The observational results complement the simulation results, finding values of $b$ spanning the full range of turbulence driving, from strongly solenoidal $b=0.24$ \citep{Federrath_2016_Brick} to strongly compressive $b\gtrsim1$ \citep{Menon_2020}, with the majority of results found in the more compressive turbulence driving regime \citep[see Fig.~12 from][]{Gerrard_2023}. The significant fluctuations in the value of $b$ and the driving mode turbulence influences the structure and dynamics of the ISM and the star formation. 

To expand our understanding of the influences of the driving mode of turbulence on the ISM and star formation, we must extend our research past the Milky Way. However, extending these analyses to extragalactic sources presents additional challenges due to lower resolution and reconstructing three-dimensional properties from two-dimensional observations, necessitating new techniques for robust measurements. To date, only two studies have measured the turbulence driving parameter outside the halo of the Milky Way. \citet{Sharda_2021} analysed the Papillon Nebula (N159E), a star-forming region in the Large Magellanic Cloud. Using \CO observations, they found $b=0.87 (\pm 25\%)$, indicating predominantly compressive turbulence driving. More recently, \citet{Gerrard_2023} studied the Small Magellanic Cloud (SMC) using GASKAP-HI observations and a novel ``roving kernel'' technique to measure $b$ spatially across the entirety of the SMC, measuring $\sigmarho$, $\mach$, and $b$ in the local kernel regions unlike the full field averaged results by \citet{Sharda_2021}. \citet{Gerrard_2023} find a median value of $b=0.51$, suggesting predominantly compressive turbulence, but local 16th-to-84th-percentile variations between $b\sim0.3$ and $b\sim 1$, and hence the driving mode of turbulence ranging through compressive to solenoidal regimes. 

Despite significant progress in understanding turbulence-driven star formation, measuring $b$ in extragalactic systems is challenging due to resolution limits, projection effects, the lack of magnetic field estimates and the overall low signal to noise in most of the samples. Hence, only two studies to date have reported values of $b$ outside the Galaxy, in the Magellanic Clouds. Extending such measurements to entire galaxies is therefore crucial for linking small-scale turbulence to galaxy-wide star formation and ISM regulation.

In this study we present the first spatially resolved measurements of turbulence driving modes across full galactic discs beyond the Milky Way, using Atacama Large Millimeter/submillimeter Array (ALMA) \CO observations of NGC7793 and NGC1313 at $\sim13\,\pc$ resolution. By applying a kernel-based analysis of density and velocity fluctuations, we map $b$ across different galactic environments, quantify both local and radial variations, and compare two morphologically distinct spirals. This study establishes a new framework in bridging small-scale cloud turbulent physics with the global galactic environment and star formation regulation. 

This paper is divided as follows. Sec.~\ref{sec:obs} describes the ALMA observational data used in this work for the galaxies NGC1313 and NGC7793. The methodology for measuring $\sigmarho$, $\mach$, and $b$ is presented in Sec.~\ref{sec:methodology}. Sec.~\ref{sec:results} presents the results, including the full field roving kernel analysis and regional variations in the turbulent statistics. We discuss potential caveats in Sec.~\ref{sec:caveats} and summarise our findings in Sec.~\ref{sec:conclusion}.

\section{Observations} \label{sec:obs}

To quantify turbulence driving in external galaxies, we require observations with sufficient spatial resolution to resolve cloud-scale density and velocity fluctuations. For this purpose, we use Atacama Large Millimeter/submillimeter Array (ALMA) CO (J=2–1) observations of the molecular gas in NGC7793 and NGC1313, which provide $13\,\pc$ spatial resolution across their full optical discs.

\subsection{Target Galaxies} 

\subsubsection{NGC7793: the flocculent galaxy}

NGC7793 (Fig.~\ref{fig:NGC7793_OptImage}) is a flocculent spiral galaxy characterised by a lack of strong spiral structure \citep{Elmegreen_1984}. It is a member of the Sculptor Group and relatively nearby at a distance of $3.62\,\mathrm{Mpc}$ \citep{Anand_2021}. NGC7793 has no bar and only a small bulge hosting a nuclear star cluster \citep{Kacharov_2018, Elmegreen_2014}. The galaxy's warped HI disc is likely due to interactions with a nearby dwarf companion, although no significant tidal effects are observed \citep{Koribalski_2018}. 

NGC7793's star formation history has been extensively studied, with \citet{Sacchi_2019}, showing an inside-out growth pattern and recent increased star formation in the Outer regions. Resolved ALMA observations have provided constraints on molecular cloud properties \citep{Grasha_2018}, which are further analysed in \citet{Finn_2024_a, Finn_2024_b}, examining the role of sub-galactic environments in shaping the molecular cloud structure \citep{Finn_2024_b}.

\subsubsection{NGC1313: the topsy-turvey galaxy} \label{sec:NGC1313_intro}
NGC1313 (Fig.~\ref{fig:NGC1313_OptImage}), often referred to as the `Topsy-Turvey' galaxy, exhibits an asymmetric spiral morphology with a notable bar structure at a distance of $4.32\,\mathrm{Mpc}$ \citep{Anand_2021}. The galaxy shows a large disturbance in its HI velocity field in the southwest region, attributed to interactions with a nearby satellite galaxy \citep{Sandage_1979, Peters_1994}. These interactions have triggered enhanced star formation while the rest of the galaxy maintains a relatively constant SFR \citep{Silva-Villa_2012, Hernandez_2022}. The molecular gas dynamics in NGC1313 are expected to be significantly influenced by these interactions, making it an ideal case study for understanding how galactic environment affects turbulence driving.

Together, these two galaxies provide a contrast between a dynamically settled disc and one undergoing strong environmental perturbations.

\subsection{ALMA Observations} 
Both galaxies were mapped in the \CO transition with ALMA, covering their full optical discs. The observations achieved a synthesised beam of $\sim0.4\arcsec$ ($\approx 13\,\pc$) and a spectral resolution of $1\,\mathrm{km}\,\mathrm{s}^{-1}$. Typical RMS sensitivities are $0.15-0.2\,\mathrm{K}$ per channel, sufficient to trace both dense clumps and diffuse emission. These data were obtained using the $12\,\mathrm{m}$ array during Cycles~4 and~5, respectively (project code 2015.1.00782.S; PI:~K.~E.~Johnson) as part of ALMA-LEGUS \citep{Grasha_2018, Finn_2024_a, Finn_2024_b}.

For NGC7793, the total on-source total integration time was $3\,\mathrm{h}$. The observations cover a total area of $3\times 2\,\mathrm{kpc}$ (or $180'' \times 114''$).
For NGC1313, observations were conducted as a mosaic of two images: the north arm, with a total integration time of $2.2\,\mathrm{h}$; and the central and southern regions, with a total integration time of $3.75\,\mathrm{h}$. 

\begin{figure*}
    \centering
    \includegraphics[width=0.97\linewidth]{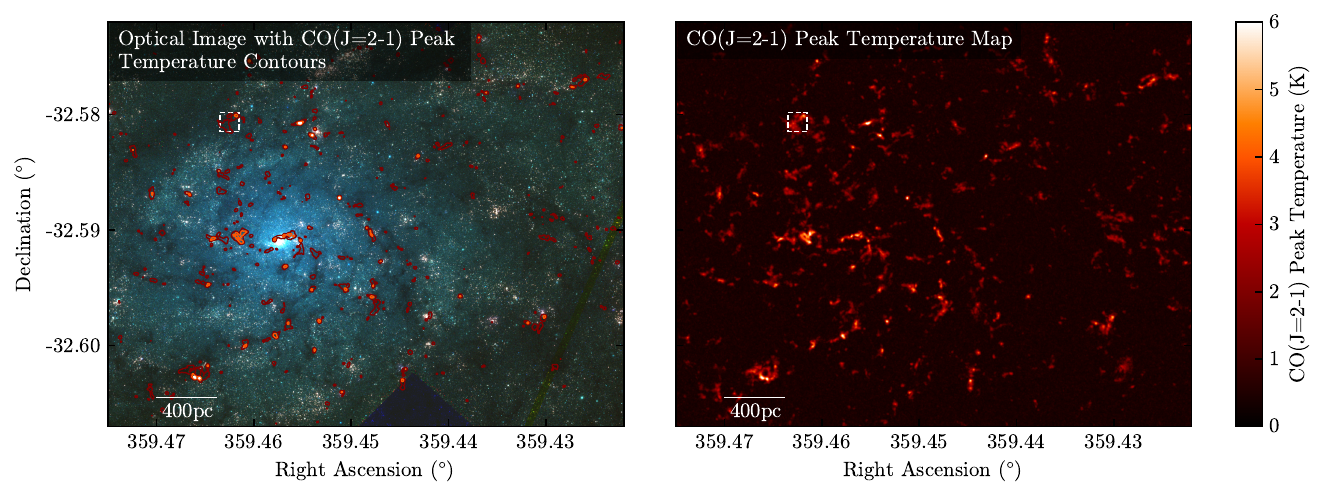}
    \caption{Optical and peak temperature ($\tpeak$) images of NGC7793. \textit{Left:} Optical image with \CO peak temperature contours overlayed. The optical image combines \textit{Hubble Space Telescope (HST)} images from the Legacy Extragalctic UV Survey (LEGUS) images \citep{LEGUS_2015}. The three-colour composite image combines the UVES/F555W (blue), UVES/F438W (green) and UVES/F255W (red). The contours are the peak temperature from the ALMA \CO emission. \textit{Right:} The peak temperature map of the \CO emission. This is the same as the contours overlaying the left-hand image with a shared colorbar. The white dashed square in both images is the subsection of the galaxy used as an example in Sec.~\ref{sec:methodology}. }
    \label{fig:NGC7793_OptImage}
\end{figure*}

\begin{figure*}
    \centering
    \includegraphics[width=\linewidth]{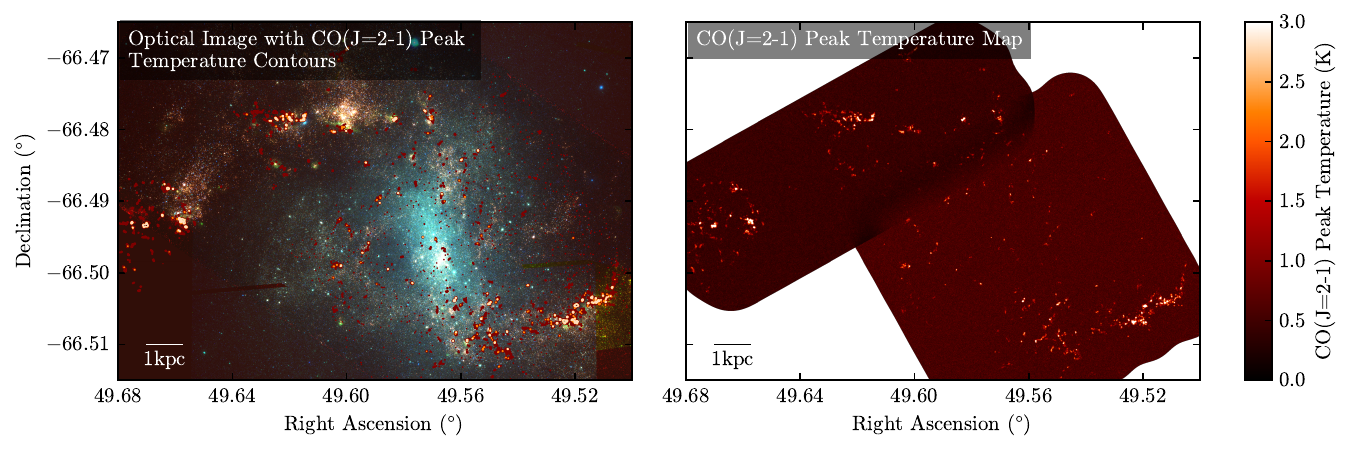}
    \caption{Same as Fig.~\ref{fig:NGC7793_OptImage}, but for NGC1313. The optical image combines \textit{HST} images from the LEGUS survey where the images are coloured such that the UVES/F814W is blue, UVES/F555W is green and UVES/F275W filter is red \citep{LEGUS_2015}.}
    \label{fig:NGC1313_OptImage}
\end{figure*}

The spatial distribution of the \CO emission across these galaxies is visualised in Fig.~\ref{fig:NGC7793_OptImage}~and~\ref{fig:NGC1313_OptImage}, with the ALMA \CO peak temperature contours overlaid on optical images from the Hubble Space Telescope (HST) Legacy Extragalactic UV Survey \citep[LEGUS; ][]{LEGUS_2015}.  

The key physical properties of both galaxies are listed in Tab.~\ref{tab:BasicInfo}. The data were calibrated using the standard ALMA pipeline in CASA. Further details of the \CO observations and data reduction are provided in \citet{Finn_2024_a}. 

\begin{table*}
    \centering
    \caption{Observational and basic galactic properties of NGC7793 and NGC1313.}
    \begin{tabular}{c c c c c c c c}
        \hline 
        Galaxy Name & \makecell{Distance $^a$ \\ (Mpc)} & \makecell{SFR $^b$ \\ (M$_\odot$ yr$^{-1}$)} & \makecell{Stellar Mass $^c$ \\ ($\times 10^9$ M$_\odot$)} & \makecell{Molecular Gas \\ Mass \\ ($\times 10^6$ M$_\odot$)} & \makecell{Beam FWHM \\ ($''$)} & \makecell{Beam FWHM \\ (pc)} & \makecell{RMS \\ (K)}  \\
        (1) & (2) & (3) & (4) & (5) & (6) & (7) & (8) \\
        \hline 
        NGC7793 & 3.62 & 0.23 & 1.78 & 14.0 & 0.72 & 13 & 0.2 \\
        NGC1313 & 4.32 & 0.68 & 1.82 & 7.6 & 0.58 & 13 & 0.15 \\ 
        \hline 
    \end{tabular}
    \begin{flushleft}
    $^a$: from \citet{Anand_2021}. \newline $^{ b \text{, } c}$: from \citet{Leroy_2019} 
    \newline Columns 5-8 list ALMA molecular gas and spectral properties from the observations as described by \citet{Finn_2024_a}.
    \end{flushleft}
    \label{tab:BasicInfo}
\end{table*}

\subsection{Signal-to-noise threshold} \label{sec:S/NThresh}

To ensure robust measurements, we apply a signal-to-noise (S/N) threshold of 5 and exclude velocity channels below this threshold. We adopt noise values of $0.2\,\mathrm{K}$ for NGC7793 and $0.15\,\mathrm{K}$ for NGC1313 \citep{Finn_2024_a}. An evaluation of the effect of S/N thresholds on turbulence parameters by \citet{Gerrard_2023} supports the use of a threshold of 5 in this study.

\subsection{Moment maps} \label{sec:MomMaps}
To determine $b$ (Equation~\ref{eqn:b_def}) we require measurements of the $3\D$ turbulent density dispersion, $\sigmarho$, and the $3\D$ turbulent Mach number, $\mach$. Integrated intensity maps provide molecular gas column densities, while velocity dispersion maps ($M_2$; second moment) trace turbulent motions. These products serve as the inputs to the turbulence analysis pipeline described in Sec.~\ref{sec:methodology}.

\subsubsection{Zeroth moment and conversion to column density} \label{sec:mom0}
The column density ($\Sigma$) is inferred from the zeroth moment ($M_0$; integrated intensity), by integration of the brightness temperature, $T_{b}(v)$, across all velocity channels ($v$) with width $dv$,
\begin{equation} \label{eqn:mom0}
    M_0 = \int T_{b}(v)\,dv \,.
\end{equation}

To convert $M_0$ to $\Sigma$ we use the CO-to-H$_2$ conversion factor, $X_\text{CO}$, from \citet[][tab.~3b]{Gong_2020}, as appropriate for the \CO line used in this study. We use a radially decreasing metallicity of $12 + \log (\text{O}/\text{H}) = 8.572-0.054\, \text{dex}\, \text{kpc}^{-1} \times R_\text{gal}$ for NGC7793 \citep{Stanghellini_2015} and a constant metallicity of $12 +\log (\text{O}/\text{H}) = 8.4 \pm 0.1$ in NGC1313 \citep[see][]{Walsh_1997}. The choices of metallicity follows the process by \citet{Finn_2024_a}, who find that the choice of literature values of $X_\text{CO}$ do not significantly affect any of the moment maps. 

\subsubsection{Second moment -- velocity dispersion}
$M_2$ provides the line-of-sight velocity dispersion, $\sigma_{v,1\D}$. The calculation of $M_2$ begins with evaluating the first moment ($M_1$), given by
\begin{equation}
    M_1 = \frac{\int v\,T_{b}(v)\,dv}{\int T_{b}(v)\,dv} = \frac{\int v\,T_{b}(v)\,dv}{M_0} \, ,
\end{equation}
which yields
\begin{equation} \label{eqn:mom2}
    M_2 = \sigma_{v,1\D} = \left(\frac{\int (v - M_1)^2\,T_{b}(v)\,dv}{M_0}\right)^{1/2} \,.
\end{equation}
$M_2$ can be used to estimate the $3\D$ turbulent velocity dispersion from $\sigma_{v,1\D}$ \citep[][as discussed in Sec.~\ref{sec:methodology}]{Stewart_2022}.

\section{Turbulence analysis methods} \label{sec:methodology}
To determine the turbulence driving parameter $b$ (Equation~\ref{eqn:b_def}), one needs to measure the $3\D$ turbulent density dispersion, $\sigmarho$, and the $3\D$ Mach number, $\mach$. Sec.~\ref{sec:MomMaps} demonstrates the process used to measure the column density, $\Sigma$, and the LOS velocity dispersion, $\sigma_{v,1\D}$, and here we explain how these quantities are converted to $3\D$ dispersions. Moreover, turbulence is a scale-dependent process \citep{Frisch1995,FederrathEtAl2021}, and hence, we now introduce the turbulent kernels in which we measure $\sigmarho$ and $\mach$.

\subsection{Turbulent kernels} \label{sec:TurbKernels}
To capture local turbulent variations in column density, $\Sigma$, and line-of-sight velocity dispersion, $\sigvod$, we apply a Gaussian kernel whose full width at half maximum (FWHM), $D_\mathrm{kernel}$, spans five beam widths, i.e., $D_\mathrm{kernel}=5\,D_\mathrm{beam}\sim65\,\pc$, given that each ALMA beam has a FWHM of $D_\mathrm{beam}\sim13\,\pc$ at the respective distance of the two galaxies (i.e., in terms of angular size, the beam FWHM is $0.72\arcsec$ in NGC7793 and $0.58\arcsec$ in NGC1313; Tab.~\ref{tab:BasicInfo}). This kernel size ensures a sufficient number of independent resolution elements for robust, kernel-weighted statistics \citep{sharda_2018}, while minimising the size in order to probe the spatial variations of the turbulent parameters.

\begin{figure*}
    \centering
    \includegraphics[width=\textwidth]{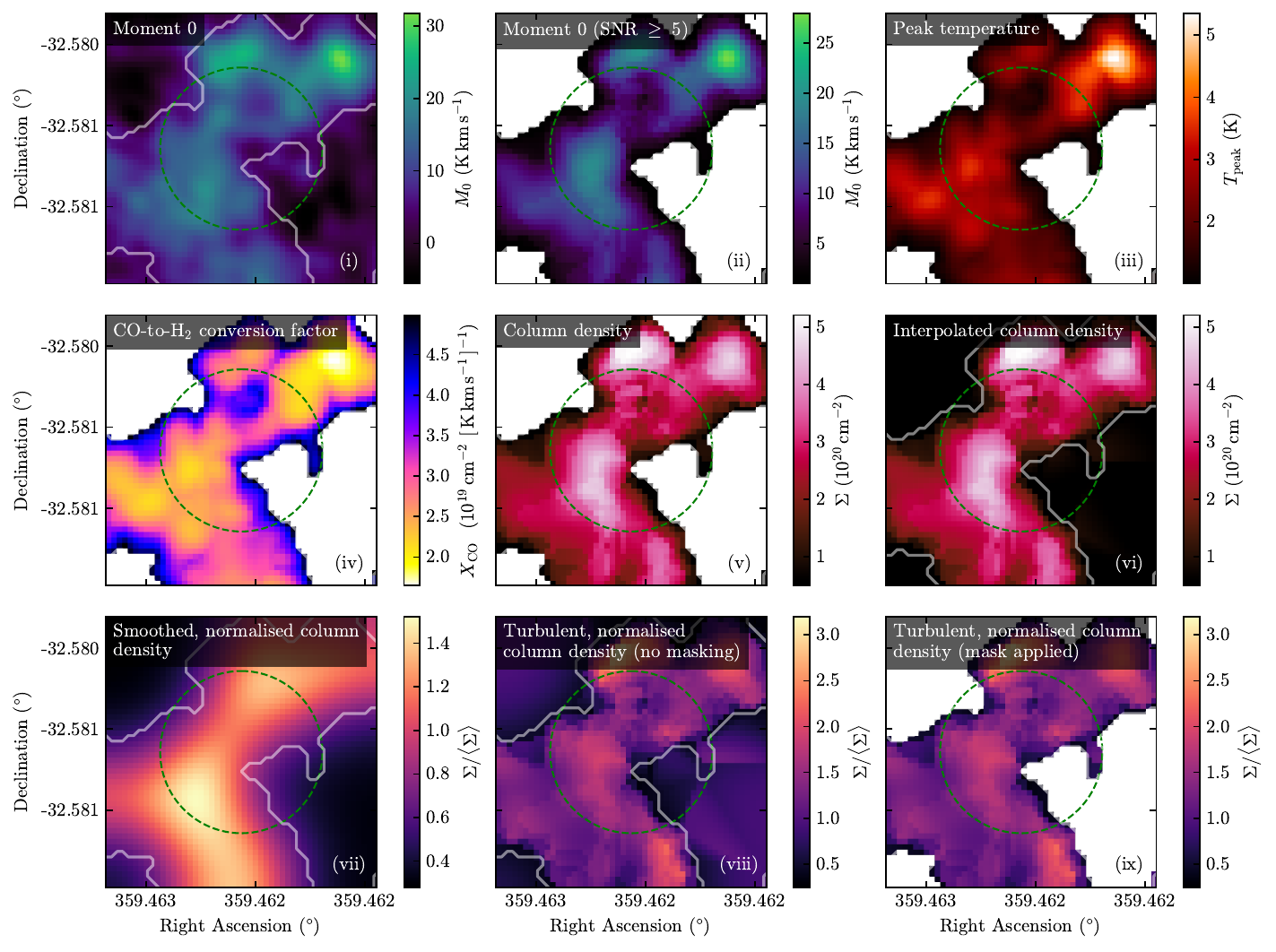}
    \caption{Example demonstration of the steps in the turbulent column density analysis for a subsection of NGC7793 (shown as the dashed square in Fig.~\ref{fig:NGC7793_OptImage}). Panels (i)--(ix) show $M_0$ without S/N threshold (i), $M_0$ with S/N threshold of 5 (ii), CO peak temperature map (iii), CO-to-H$_2$ conversion factor, $X_\text{CO}$ (iv), column density, $\Sigma=X_\text{CO} M_0$ (v), linearly interpolated column density (vi), mean-normalised and Gaussian-smoothed column density (vii), turbulence-isolated column density (viii), and the same as (viii), but with the S/N mask reapplied (ix). The dashed circle indicates the kernel FWHM of 5~beam widths ($65\,\pc$ in diameter).}
    \label{fig:Mom0_kernel_subtr}
\end{figure*}
We note that the size of the kernel influences the value of the components that determine $b$, which is explored thoroughly in \citet{Gerrard_2023}, who showed that whilst the kernel size has an impact on $\sigmarho$ and $\mach$ individually, the $b$-parameter does not significantly depend on the kernel size, provided it is large enough to encompass multiple beam widths, as the changes in $\sigmarho$ with kernel size are roughly proportional to the changes in $\mach$.

For the following kernel-based analysis, we compute the kernel-weighted statistics (mean and standard deviation) of $\Sigma$ and $\sigvod$. The kernel-weighted mean of any quantity $q$ is defined as
\begin{equation}
\label{eqn:kernweightedmean}
\langle q \rangle = \frac{\int q\,W(r)\,dr}{\int W(r)\, dr},
\end{equation}
with the kernel function, $W(r)$, given by $W(r)=W_\text{un-norm}(r) / \sum W_\text{un-norm}(r)$ and
\begin{equation} \label{eqn:weightingfunct}
    W_\text{un-norm}(x,y) = \exp \left( \frac{-((x-x_0)^2+(y-y_0)^2)}{2\sigma_\mathrm{kernel}^2} \right) ,
\end{equation}
with $\sigma_\mathrm{kernel}=D_\mathrm{kernel}/(2\sqrt{2\ln 2)}$, and $x$ and $y$ are found depending on the radius from the central pixel, $r$. In practise, Equations~\ref{eqn:kernweightedmean}~and~\ref{eqn:weightingfunct} are computed as finite sums over all pixels with $r\leq3\,\sigma_\mathrm{kernel}$, i.e., the kernel is numerically truncated at a sufficiently large distance $r$, such that the truncation does not introduce any significant errors. Pixels with no velocity channels above a S/N of 5 (as defined in Sec.~\ref{sec:S/NThresh}) are assigned zero weight. We define the kernel‐weighted coverage as
\begin{equation}
  P \;=\;\frac{\displaystyle\int_{\rm valid}W(r)\,dr}{\displaystyle\int_{\rm kernel}W(r)\,dr}\approx\frac{\sum_{\rm valid}W(r)}{1}\,,
\end{equation}
where $W(r)=0$ at any spaxel where no velocity channels reach the S/N threshold in the valid integral so that ${0\le P\le1}$.  We require $P\ge0.5$ for a majority of valid data; kernels with $P<0.5$ are omitted and their central pixels left blank.

The corresponding kernel-weighted standard deviation of $q$ is calculated as
\begin{equation} \label{eqn:stdev}
\sigma_q = \sqrt{\langle q^2 \rangle - \langle q \rangle^2}.
\end{equation}
where the $\langle \rangle$ operator corresponds to a kernel-weighted average, Equation~\ref{eqn:kernweightedmean}. 

\subsection{Measuring the turbulent, volumetric density dispersion} \label{sec:densitydisp}

To measure the turbulent, volumetric density dispersion, $\sigmarho$, we first compute the column density dispersion, and then convert this to the volume-density dispersion.

\subsubsection{Interpolation to fill low-S/N spaxels} \label{sec:interp}

The conversion from column-density ($\Sigma$) to volume-density ($\rho$) dispersion requires an estimate of the power spectrum of the column density in each kernel (as discussed in detail in Sec.~\ref{sec:vol_disps} below). Since calculating the power spectrum involves a Fourier transformation, we require a filled map of the column density, where missing information (spaxels below the S/N threshold) is obtained by linear interpolation with \texttt{scipy.griddata}. We discuss the impact of the interpolation in Appendix~\ref{sec:bruntR_interpimpacts}. The same interpolation also allows for the Gaussian smoothing process required to separate turbulence from non-turbulent fluctuations.

\subsubsection{Turbulence isolation} \label{sec:LowPassFilter}

There are systematic, large-scale, non-turbulent variations in the density and velocity fields of a galaxy. For example, the surface density is usually higher in the centre compared to the arms or interarm regions. Such large-scale variations are not considered turbulence and need to be separated from the turbulent fluctuations that we aim to measure. Thus, we isolate the turbulence by creating a smoothed (low-pass filtered) column-density map and subtract the smoothed map from the original map, leaving only the turbulent density fluctuations. 

The smoothed column density is created by using a Gaussian filter. For the Gaussian smoothing we use the \texttt{scipy.ndimage.gaussian\_filter} package. Following \citet{Gerrard_2024}, we set the FWHM of the smoothing kernel to half the FWHM of the roving kernel (2.5~beam widths, or $\sim 33\,\mathrm{pc}$). By applying this method to the logarithm of the interpolated column density, we capture the large-scale density fluctuations \citep{Gerrard_2024}. From this we determine the standard deviation of the normalised column density, $\sigma_{\SigmaN}$, by Equation~\ref{eqn:stdev}. We note that measurement of $\sigma_{\SigmaN}$ does not include any of the interpolated spaxels, only those with valid data before the interpolation process.

\subsubsection{Converting column density dispersion to volumetric density dispersion} \label{sec:vol_disps}
To convert the derived column density dispersion into the volumetric density dispersion required in Equation~\ref{eqn:b_def}, we use the method in \citet{Brunt_2010}. This method utilises the power spectrum of the column density and extends it into $3\D$ space. The $3\D$ power spectrum, $P_{3\D}(k)$ can be found as per the relationship given by \citet{Federrath_2013}, 
\begin{equation} \label{eqn:3D_2D_powerspect}
    P_{3\D}(k) = 2k\,P_{2\D}(k),
\end{equation}
where $k$ is the wave number and $P_{2\D}(k)$ is the $2\D$ power spectrum, which is computed by Fourier transformation of the turbulence-isolated column density as per Sec.~\ref{sec:LowPassFilter}. 
Finally, we use the relationship provided by \citet{Brunt_2010},
\begin{equation} \label{eqn:Brunt_R}
    \mathcal{R}^{1/2} = \frac{\sigma_{\Sigma / \langle \Sigma \rangle}}{\sigmarho} = \left(\frac{\int P_{2\D}(k)\,dk}{\int P_{3\D}(k)\,dk}\right)^{1/2},
\end{equation}
to obtain $\sigmarho$, where the integrals are evaluated as discretised summation over $k$-space in our implementation. Equations~\ref{eqn:3D_2D_powerspect} and \ref{eqn:Brunt_R} make the assumption that the density power spectrum is isotropic \citep[see][]{Brunt_2010,Federrath_2013,Federrath_2016_Brick,Gerrard_2023}, and we discuss this assumption in Sec.~\ref{sec:IsoDensPowerField} and Appendix~\ref{sec:PowerSpectEG}. Since the turbulent kernels are non-periodic, we apply mirroring of the kernel data before computing the Fourier transform to minimise artefacts in the power spectra.

\subsection{Demonstration of kernel-based turbulent density estimate}
Here we summarise the methods introduced so far for deriving the turbulent density statistics based on an example kernel. Fig.~\ref{fig:Mom0_kernel_subtr} illustrates the first steps in the method of deriving $\sigmarho$ for one subsection of NGC7793. This subsection is outlined by a white square in Fig.~\ref{fig:NGC7793_OptImage}. The green, dashed circle in all of the nine panels of Fig.~\ref{fig:Mom0_kernel_subtr} shows the FWHM of the turbulent kernel (see Sec.~\ref{sec:TurbKernels}). Initially, we apply a S/N threshold of 5 in all velocity channels and compute $M_0$ as per Sec.~\ref{sec:MomMaps}. The effect of applying the S/N threshold is illustrated in panels (i) and (ii) of Fig.~\ref{fig:Mom0_kernel_subtr}. The S/N threshold leaves some spaxels without any data to determine the $M_0$ from, which are left as blank pixels in panels (ii) to (v). As per Sec.~\ref{sec:mom0}, we use $\tpeak$ (panel~(iii)) to determine $X_\text{CO}$ (panel~(iv)). The product of $M_0$ and $X_\text{CO}$ gives the column density, $\Sigma$ (panel~(v)). Interpolating this (see Sec.~\ref{sec:interp}) fills the empty spaxels with buffer data (panel~(vi)) in preparation for the turbulence-isolation method. As per Sec.~\ref{sec:LowPassFilter} the column density is normalised by the kernel-weighted mean, and is smoothed with a Gaussian kernel (panel~(vii)). This smoothed map is subtracted off the normalised column density to yield the turbulent column density (panel~(viii)), and finally the mask of the pixels where no velocity channels reach a S/N of 5 is reapplied (panel~(ix)). The turbulent column density fluctuations are computed from this final map.

\begin{figure}
    \centering
    \includegraphics[width=\linewidth]{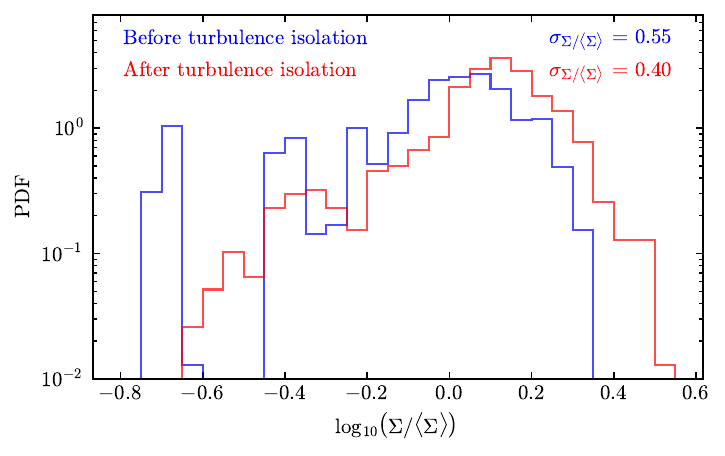}
    \caption{PDFs of the logarithm of the normalised column density, $\SigmaN$ (blue histogram) compared to the turbulence-isolated version of $\SigmaN$ (red histogram) of the kernel section shown in Fig.~\ref{fig:Mom0_kernel_subtr}. The shape of the distribution becomes more Gaussian and the standard deviation is reduced after turbulence isolation, as the non-turbulent column density fluctuations are removed by the turbulence-isolation procedure (see Sec.~\ref{sec:LowPassFilter}).}
    \label{fig:PDFs}
\end{figure}

The efficacy of the turbulence isolation is quantified in the change of shape of the PDF of $\log_{10}(\SigmaN)$, which is shown in Fig.~\ref{fig:PDFs}. The PDF of the original and turbulence-isolated column density are shown in Fig.~\ref{fig:PDFs}, in blue and red, respectively. We see strongly non-Gaussian features in the PDF before turbulence isolation (e.g., the separated low-density peak). In contrast, after turbulence isolation, the PDF is closer to a Gaussian distribution, as expected for a turbulent medium \citep{Passot_1998,Kritsuk_2007,Federrath_2008,Federrath_2010}. Furthermore, the standard deviation of the turbulence-isolated data is reduced compared to the original data, as the large-scale, systematic, non-turbulent fluctuations have been subtracted out \citep[see also][]{Stewart_2022}.

The $\sigma_{\SigmaN}$ after turbulence isolation is then converted to the volumetric standard deviation, $\sigmarho$, through the process outlined in Sec.~\ref{sec:vol_disps}.

\subsection{Measuring the turbulent Mach number} \label{sec:Mach}

The turbulent Mach number, $\mach$, is the ratio of the $3\D$ turbulent velocity dispersion to the sound speed, $\mach=\sigvtd/\cs$.

\subsubsection{Converting $\sigvod$ to $\sigvtd$} \label{sec:sigmav1d_3d}

Based on $\sigvod$ from $M_2$ (Equation~\ref{eqn:mom2}), we obtain the $3\D$ turbulent velocity dispersion as
\begin{equation} \label{eq:sigvtd}
\sigvtd = \sigvod\times\mathcal{C}_\mathrm{i}^\mathrm{any},
\end{equation}
with the $M_2$ conversion factor $\mathcal{C}_\mathrm{i}^\mathrm{any}$ from \citet[][tab.~E1, rows 10--12]{Stewart_2022}. Those three entries in tab.~E1 of \citet{Stewart_2022} correspond to three different ratios of the turbulent-to-rotational energy (ratios of 0, 0.35, and 1). We average across these three entries, taking into account the uncertainties in each case reported in \citet{Stewart_2022}, which results in $\mathcal{C}_\mathrm{i}^\mathrm{any}=(-0.97 \pm 0.57) (f/R) + 2.34\pm0.33 = 2.0 \pm0.8$, where $f/R$ is the ratio of the beam FWHM to the kernel radius, i.e., $f/R=D_\mathrm{beam}/(D_\mathrm{kernel}/2)=1/(5/2)=0.4$, given the kernel definition from Sec.~\ref{sec:TurbKernels}. We note that $M_2$ does not allow for the same turbulence-isolation process as for the column density. Thus, non-turbulent contributions to $M_2$ may somewhat artificially increase the value of $\mach$ derived based on $M_2$, but \citet{Stewart_2022} show that the effect is typically $\ll50\%$ \citep[see top right panel in Fig.~3 of][]{Stewart_2022}. We note that the method outlined by \citet{Stewart_2022} assumes spherical geometry. This is a reasonable approximation for our purposes, as we explicitly use small-scale, spherically symmetric Gaussian kernels in the roving-kernel approach described below (Sec.~\ref{sec:RovKern}).

\subsubsection{Sound speed ($\cs$)} \label{sec:soundspeed}
The sound speed is defined as
\begin{equation} \label{eqn:soundspeed}
    \cs = \sqrt{\frac{k_\text{B}T}{\mu_\text{p}m_\text{H}}},
\end{equation}
where $k_\text{B}$ is the Boltzmann constant, $T$ is the gas temperature, $\mu_\text{p}$ is the mean particle weight, and $m_\text{H}$ is the mass of a hydrogen atom. 

Although our observations trace molecular gas via the \CO line, the sound speed and gas kinematics on the kernel scale may be significantly influenced by contributions of the cold neutral medium (CNM). The CNM and molecular gas differ in temperature and mean particle weight. At solar metallicity, the CNM typically spans $\sim50-200\,\mathrm{K}$ with $\mu_\text{p}\sim1.3$ \citep{McClure_Griffiths_2023}, while molecular gas has $\sim10-50\,\mathrm{K}$ with $\mu_\text{p}\sim2.3$ \citep{Omont_2007}. Applying Equation~\ref{eqn:soundspeed}, we approximate an average sound speed of $\sim 0.33\,\mathrm{km\,s^{-1}}$ and $0.89\,\mathrm{km\,s^{-1}}$ for pure molecular and CNM gas, respectively. Assuming a multiphase ISM which is evenly split between the CNM and molecular gas, we adopt $\cs=0.6\pm0.3\,\mathrm{km\,s^{-1}}$, taken as the arithmetic mean of the two $\cs$ values. We discuss the influence of this approximation in more detail in Sec.~\ref{sec:caveats}.

\begin{table*}
    \centering
    \caption{Uncertainty propagation of all individual kernel measurements}
    \renewcommand{\arraystretch}{1.4}
    \begin{tabular}{cccccccc} 
         \hline
         Galaxy & $\sigma_{\SigmaN}$ & $\mathcal{R}^{1/2}$ & $\sigmarho$ & $\sigma_{v,1\D}$ & $\sigma_{v, {3\D}}$ & $\mach$ & $b$ \\
         \tiny{(1)} & \tiny{(2)} & \tiny{(3)} & \tiny{(4)} & \tiny{(5)} & \tiny{(6)} & \tiny{(7)} & \tiny{(8)} \\
         \hline 
         NGC7793 &17\% & $20$\% & 26\% & 24\% & 48\% & 69\% & 74\% \\
         NGC1313 & 21\% & $ 20$\% & 29\% & 13\% & 43\% & 66\% & 72\% \\
         \hline
    \end{tabular}
    \begin{flushleft}
    Notes: All relative uncertainties for individual kernel measurements are propagated by the process discussed in Sec.~\ref{sec:uncertainty_prop}. The relative uncertainties are derived for col.~(2) and~(5) from the observational uncertainties \citep{Finn_2024_a}; for col.~(3) from \citet{Brunt_2010}; for col.~(4) by error propagation of col.~(2) and~(3); for col.~(6) from the conversion factor of the 1D-to-3D turbulent velocity dispersion, $\mathcal{C}_\mathrm{i}^\mathrm{any}=1.95\pm 0.8$ \citep[][tab.~E1]{Stewart_2022}; for col.~(7) from the uncertainty in the sound speed, $\cs=0.6\pm0.3\,\mathrm{km}\,\mathrm{s}^{-1}$, i.e., a $50\%$ relatively uncertainty propagated with the uncertainty in col.~(6); and for col.~(8) by error propagation of col.~(4) and~(7).
    \end{flushleft}
    \label{tab:uncertaintyProp}
\end{table*}

\subsection{Roving the kernel} \label{sec:RovKern}
For each individual kernel defined in Sec.~\ref{sec:TurbKernels}, we compute $\sigmarho$ (see Sec.~\ref{sec:densitydisp}) and $\mach$ (see Sec.~\ref{sec:Mach}). A single value for each of the turbulent parameters $b$, $\mach$, and $\sigmarho$ is then assigned to the central pixel of the kernel. 

Following a method similar to \citet{Gerrard_2023}, we systematically move the kernel across the galaxy one spaxel at a time --- a process we refer to as the `roving kernel'. Not all kernels are evaluated; many are excluded if they do not reach a kernel-weighted majority of valid data (see Sec.~\ref{sec:TurbKernels}). We find that after the S/N threshold of 5 is introduced, $56\%$ of the total signal remains in NGC7793, and $94\%$ in NGC1313. Following the additional kernel selections described in Sec.\ref{sec:TurbKernels}, $23\%$ of the total signal remains in NGC7793 and $39\%$ in NGC1313, reflecting the stringent requirement for a sufficient fraction of valid data points per kernel.

\subsection{Uncertainty calculations} \label{sec:uncertainty_prop}

The uncertainties of each measurement have been precisely evaluated by first evaluating the kernel uncertainties and then propagating these through to the uncertainties of the larger, regional measurements. 

The kernel measurement uncertainties of each value measured in this process are shown in Tab.~\ref{tab:uncertaintyProp}, with the origin of each uncertainty detailed in the table notes. Standard error propagation is used for all derived quantities.

The uncertainties of each kernel measurement are then propagated to the median values of each regional measurement, which combines multiple kernels depending on the region size. The regions are defined later (Sec.~\ref{sec:LocVar}), but correspond to large-scale, distinct galactic morphological sectors, such as spiral arms in NGC1313 and radial rings in NGC7793. By stacking multiple kernel measurements we can quantify the uncertainty of the median. For any measurement quantity $q$ the uncertainty of each region listed in Tab.~\ref{tab:RegionalVarDependencies} is computed as 
\begin{equation}
    \left[\frac{\sigma_q}{q}\right]_{\text{region}} = \sqrt{\frac{\pi}{2}}\frac{1}{\sqrt{n_\text{kernels} - 1}} \left[\frac{\sigma_q}{q}\right]_{\text{indiv.}} \, ,
\end{equation}
where $\left[\sigma_q / q\right]_{\text{region}}$ is the relative uncertainty of the median of $q$ for the region, $\left[\sigma_q / q\right]_{\text{indiv.}}$ is the relative uncertainty of an individual kernel measurement (see Tab.~\ref{tab:uncertaintyProp}), and $n_\text{kernel}$ is the number of kernels in the region.

The factor $\sqrt{\pi/2}$ accounts for the ratio between the standard error of the median and the standard error of the mean for normally distributed data \citep[e.g.,][]{casella_2002, Maindonald_Braun_2010}. This correction accounts for the broader expected scatter of the median compared to the mean. The resulting regional uncertainties of all median values are listed in Tab.~\ref{tab:RegionalVarDependencies}. 

\section{Results} \label{sec:results}

Here we present the main results of the turbulence analysis of the galaxies NGC7793 and NGC1313. We first discuss the full fields and then focus on the local and radial spatial variations within each galaxy. Quantitative measurements are presented in the format $q = {q_{50}(\pm\sigma_q)}^{q_{84}-q_{50}} _{q_{16}-q_{50}}$, where $q$ is a measured quantity (e.g., $\mach$, $\sigmarho$, and $b$), with $q_{50}$, $q_{16}$, and $q_{84}$ being the 50th (median), 16th, and 84th percentile values, respectively, representing the spatial variations and $\sigma_q$ is the measured uncertainty of $q_{50}$. The uncertainty of the medians of each region are shown following the measurement in Tab.~\ref{tab:RegionalVarDependencies}, as calculated in Sec.~\ref{sec:uncertainty_prop}. 

\begin{figure*}
    \centering
    \includegraphics[width=0.96\linewidth]{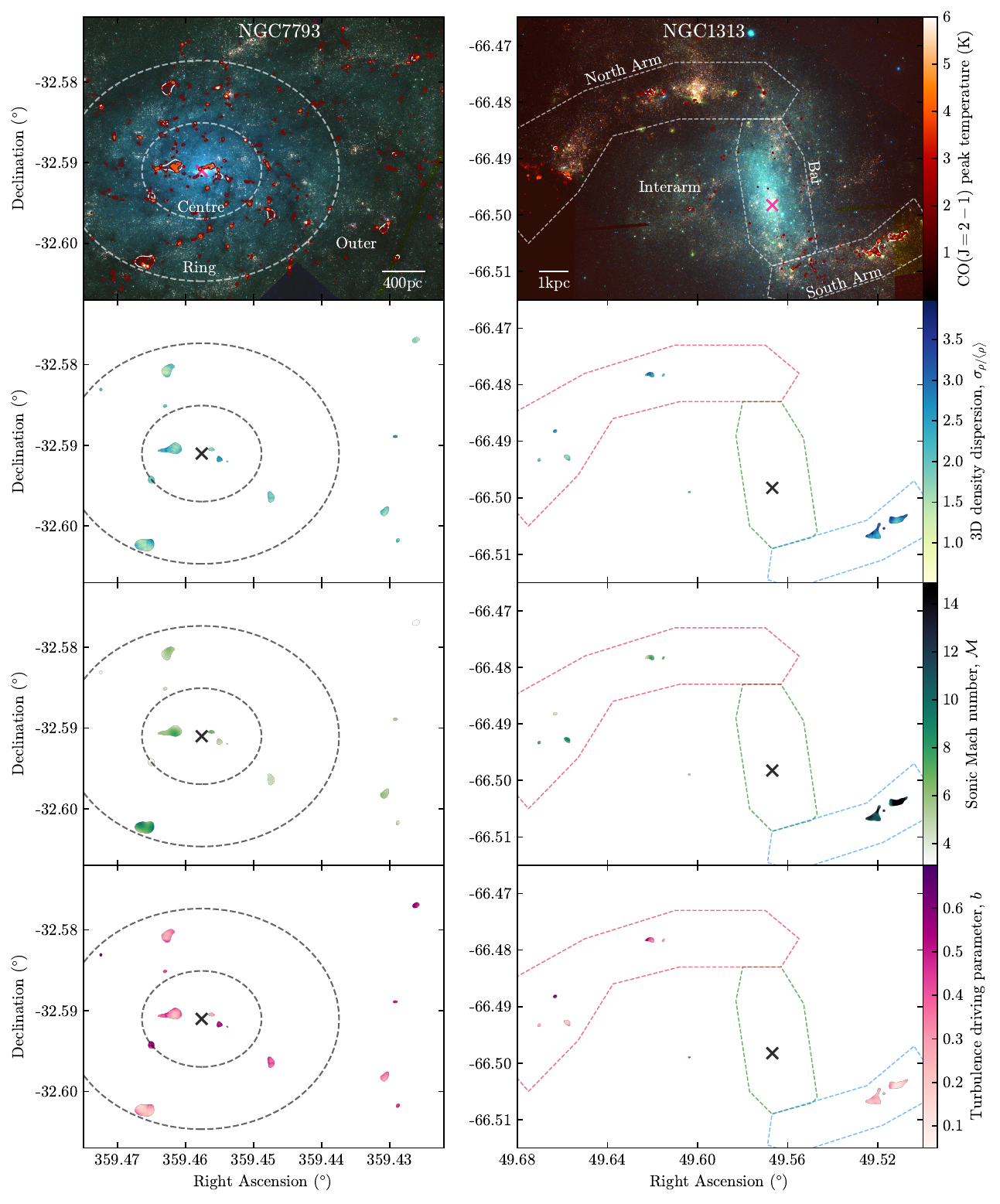}
    \caption{Maps of the \CO peak temperature, $\tpeak$, as in Figs.~\ref{fig:NGC7793_OptImage} and~\ref{fig:NGC1313_OptImage} (1st row), turbulent density dispersion, $\sigmarho$ (2nd row), Mach number, $\mach$ (3rd row), and turbulence driving parameter, $b$ in NGC7793 (left) and NGC1313 (right). The white contours in the top panels outline regions in which the turbulence analysis yielded valid results based on several criteria (see Sec.~\ref{sec:RovKern}). The dashed contours define regions defined in \citet{Finn_2024_b}, as annotated: the Centre, Ring and Outer regions in NGC7793, and the North Arm, Bar, South Arm, and Interarm regions in NGC1313. The galaxy centres are marked with a cross. While the turbulence analysis can only be successfully performed in a few local regions (hence the sparseness of the turbulence data), they span across the entire field of view of both galaxies, and show substantial spatial variation.}
    \label{fig:NGC7793_NGC1313_b_field}
\end{figure*}

Fig.~\ref{fig:NGC7793_NGC1313_b_field} shows the main result of the turbulence analysis, presenting the full-field maps of (from top to bottom) $\tpeak$, $\sigmarho$, $\mach$, and $b$ in NGC7793 (left) and NGC1313 (right). In the following, we discuss global and local properties in detail, so we refer back to this figure throughout the following discussion.

\subsection{Full field} \label{sec:FullField}

\begin{figure*}
    \centering
    \includegraphics[width=0.98\linewidth]{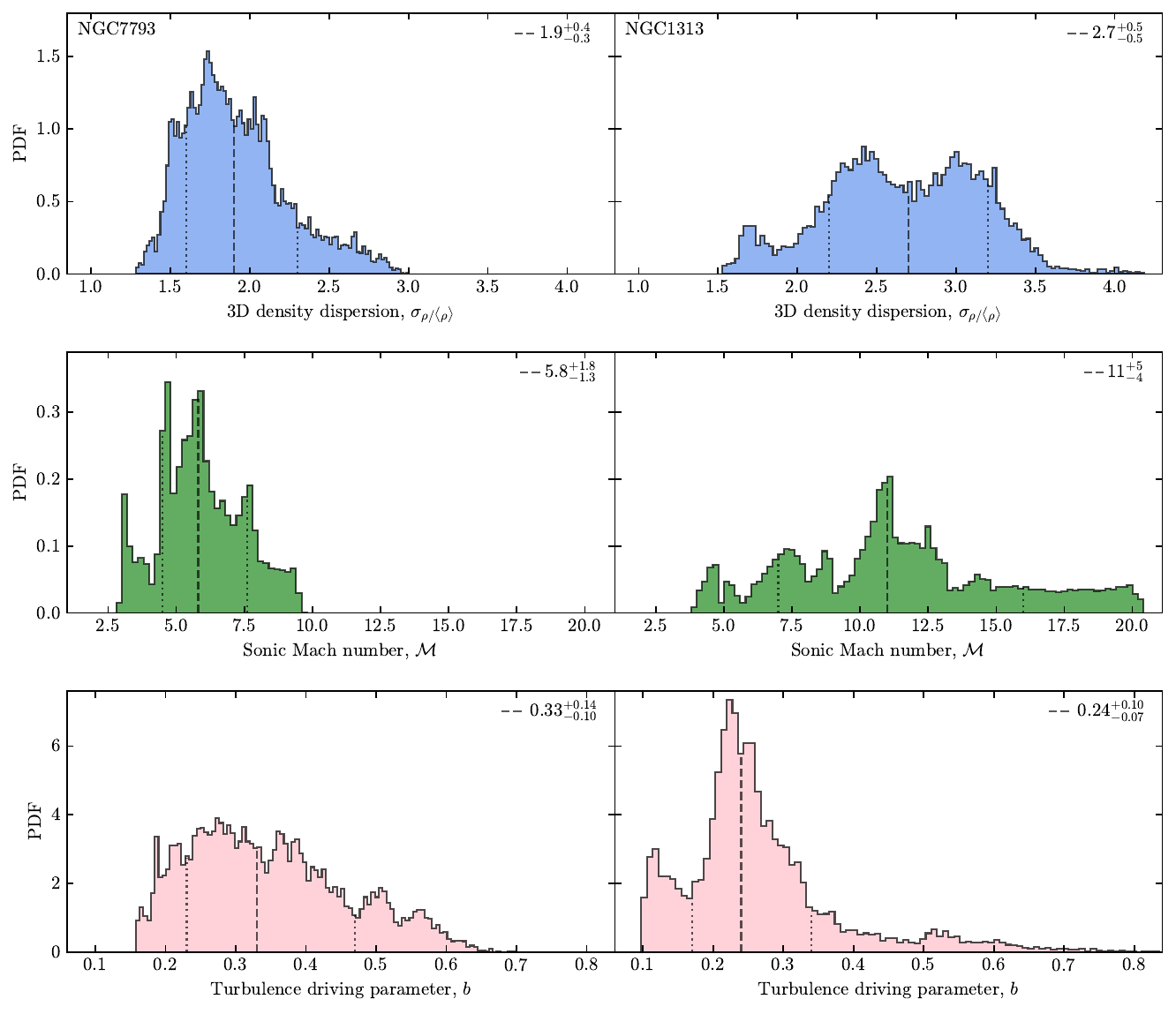}
    \caption{Full-field PDFs of $\sigmarho$ (top), $\mach$ (middle), and $b$ (bottom) for NGC7793 (left) and NGC1313 (right). The median (50th), and the 16th and 84th percentile values of the PDFs are shown as the dashed and dotted lines surrounding the median, respectively, with the values quoted in the panel legends. These PDFs quantify the averages and the spatial variations of the turbulence parameters within each galaxy, which we find are significant. In particular, the median turbulence driving mode $b\sim0.33$ (NGC7793) and $0.24$ (NGC1313) suggests primarily solenoidal driving of turbulence in both galaxies, however, with some regions reaching into the compressive-driving regime ($b>0.4$).}
    \label{fig:NGC7793_NGC1313_full_field_PDFs}
\end{figure*}

\subsubsection{Full-field maps}

Fig.~\ref{fig:NGC7793_NGC1313_b_field} shows the full-field maps of NGC7793 and NGC1313. Despite the stringent S/N cuts (Sec.~\ref{sec:S/NThresh}) and the other kernel requirements (Sec.~\ref{sec:TurbKernels}) we still find significant measurements across the fields of the two galaxies. Valid measurements align closely with regions of high peak temperature in the optical images, confirming our criteria effectively exclude low statistical significance data. In NGC1313 the regions of valid data generally follow the spiral arms, where we expect the majority of the \CO emission. Thus, it is not surprising that this is where we find the majority of results. Despite the sparseness of valid turbulence measurements, the data spans across different regions of the galaxies, allowing us to probe the various environments as well as determine galaxy-wide properties for the regions where we find valid results. Predominant of the galaxy-wide properties is the spatial variance of the turbulent parameters. There are clear signatures of $\gtrsim2$ spatial variation in all parameters, which we quantify further below. 

\begin{table*}
    \caption{Turbulence statistics in different regions of NGC7793 and NGC1313.}
    \renewcommand{\arraystretch}{1.5}
    \setlength{\tabcolsep}{2.4pt}
    \begin{tabular}{llccccccc} 
         \hline
         Region & $n_\text{kernels}$ & $\sigma_{\SigmaN}$ & $\mathcal{R}^{1/2}$ & $\sigmarho$ & $\sigma_{v,1\D}$ & $\sigma_{v, {3\D}}$ & $\mach$ & $b$ \\
         \hline 
         \multicolumn{9}{c}{NGC7793} \\
         Full field & $40.4$ &$0.52(\pm0.02)^{+0.10}_{-0.06}$ & $0.28(\pm0.01)^{+0.02}_{-0.02}$ & $1.9(\pm0.1)^{+0.4}_{-0.3}$ & $1.8(\pm0.1)^{+0.6}_{-0.4}$ & $3.5(\pm0.3)^{+1.1}_{-0.8}$ & $5.8(\pm0.8)^{+1.8}_{-1.3}$ & $0.33(\pm0.05)^{+0.14}_{-0.10}$ \\
         Centre & $13.6$ & $0.54(\pm0.03)^{+0.12}_{-0.05}$ & $0.28(\pm0.02)^{+0.02}_{-0.01}$ & $2.0(\pm0.2)^{+0.4}_{-0.2}$ & $1.8(\pm0.2)^{+0.4}_{-0.5}$ & $3.5(\pm0.6)^{+0.7}_{-1.0}$ & $5.8(\pm1.4)^{+1.2}_{-1.7}$ & $0.37(\pm0.10)^{+0.14}_{-0.12}$\\
         Ring & $14.6$ & $0.50(\pm0.03)^{+0.08}_{-0.05}$ & $0.28(\pm0.02)^{+0.02}_{-0.02}$ & $1.8(\pm0.2)^{+0.4}_{-0.2}$ & $1.9(\pm0.2)^{+0.6}_{-0.5}$ & $3.8(\pm0.6)^{+1.2}_{-1.0}$ & $6.3(\pm1.5)^{+2.0}_{-1.6}$ & $0.29(\pm0.07)^{+0.11}_{-0.08}$ \\
         Outer & $12.3$& $0.55(\pm0.03)^{+0.08}_{-0.07}$ & $0.29(\pm0.02)^{+0.02}_{-0.02}$ & $1.8(\pm0.2)^{+0.4}_{-0.2}$ & $1.6(\pm0.1)^{+0.2}_{-0.6}$ & $3.1(\pm0.6)^{+0.3}_{-1.3}$ & $5.2(\pm1.3)^{+0.6}_{-2.1}$ & $0.41(\pm0.11)^{+0.12}_{-0.11}$ \\ 
         \hline 
         \multicolumn{9}{c}{NGC1313} \\
         Full field &$43.5$& $0.73(\pm0.03)^{+0.12}_{-0.12}$ & $0.27(\pm0.01)^{+0.02}_{-0.01}$ & $2.7(\pm0.2)^{+0.5}_{-0.5}$ & $3.4(\pm0.1)^{+1.4}_{-1.2}$ & $7(\pm1)^{+3}_{-2}$ & $11(\pm1)^{+5}_{-4}$ & $0.24(\pm0.03)^{+0.10}_{-0.07}$ \\
         North Arm &$13.6$& $0.66(\pm0.05)^{+0.10}_{-0.16}$ & $0.28(\pm0.02)^{+0.02}_{-0.02}$ & $2.4(\pm0.2)^{+0.5}_{-0.6}$ & $2.2(\pm0.1)^{+0.4}_{-0.7}$ & $4.3(\pm0.7)^{+0.7}_{-1.4}$ & $7.1(\pm1.7)^{+1.2}_{-2.4}$ & $0.33(\pm0.08)^{+0.20}_{-0.12}$ \\
         South Arm & $21.4$& $0.77(\pm0.04)^{+0.09}_{-0.13}$ & $0.27(\pm0.01)^{+0.01}_{-0.01}$ & $2.8(\pm0.2)^{+0.4}_{-0.5}$ & $3.8(\pm0.1)^{+1.5}_{-0.5}$ & $7.4(\pm0.9)^{+2.9}_{-1.1}$ & $12(\pm2)^{+5}_{-2}$ & $0.23(\pm0.05)^{+0.05}_{-0.08}$ \\
         Interarm & $8.5$ & $0.56(\pm0.05)^{+0.03}_{-0.02}$ & $0.24(\pm0.02)^{+0.02}_{-0.02}$ & $2.3(\pm0.3)^{+0.3}_{-0.3}$ & $1.6(\pm0.1)^{+0.1}_{-0.1}$ & $3.1(\pm0.6)^{+0.1}_{-0.1}$ & $5.2(\pm1.6)^{+0.2}_{-0.1}$ & $0.44(\pm0.14)^{+0.06}_{-0.04}$ \\
         \hline
    \end{tabular}
    \begin{flushleft}
     The values quoted are structured such that the first number is the median value for the specified region, the second number inside the brackets is the uncertainty of the median (see Sec.~\ref{sec:uncertainty_prop}) and the sub- and super-script values represent the 16th and 84th percentile spatial variation in the region.
    \end{flushleft}
    \label{tab:RegionalVarDependencies}
\end{table*}

\subsubsection{Full-field PDFs}
Fig.~\ref{fig:NGC7793_NGC1313_full_field_PDFs} shows the full-field PDFs, which quantify the averages and spatial variations of the turbulent parameters within each galaxy. The most important statistical quantities from the PDFs are listed in Tab.~\ref{tab:RegionalVarDependencies}, noting that we also include local (separated in regions) measurements, which we refer to later. We find distributions with significant spreads. The PDFs of $\sigmarho$ are significantly different between both galaxies, where $\sigmarho = 1.9(\pm0.1)^{+0.4}_{-0.3}$ in NGC7793 and $\sigmarho = 2.7(\pm0.2)^{+0.5}_{-0.5}$ in NGC1313, i.e., the latter shows a $\sim40\%$ higher level of density fluctuations. This could be explained by the higher starbursting environment in NGC1313, resulting in a higher intensity of turbulence, potentially caused by NGC1313's satellite interaction. We find a skewed-Gaussian distribution of $\sigmarho$ with an extended high-value tail in NGC7793, while NGC1313 displays a bimodal or potentially trimodal distribution.

Furthermore, we measure $\mach = 5.8(\pm0.8)^{+1.8}_{-1.3}$ in NGC7793 and $\mach=11(\pm1)^{+5}_{-4}$ in NGC1313, showing up to $\sim50\%$ variations in $\mach$ across the different regions. The PDFs of $\mach$ show multiple peaks in both galaxies, likely due to distinct turbulent environments within the galaxies. We expect a Gaussian distribution of $\mach$ within an individual turbulent environment \citep[e.g.,][]{Gerrard_2023}, and we interpret the distinct peaks in the distributions of $\mach$ as a result of the superposition of multiple Gaussian distributions of $\mach$. We find both $\sigmarho$ and $\mach$ have larger median values in NGC1313 than in NGC7793, suggesting a higher intensity of turbulence in NGC1313.

The PDFs of $b$ are shown in the bottom panels of Fig.~\ref{fig:NGC7793_NGC1313_full_field_PDFs}. With $b\geq0.33(\pm0.05)^{+0.14}_{-0.10}$ and $b\geq0.24(\pm0.03)^{+0.10}_{-0.07}$ in NGC7793 and NGC1313, respectively, the turbulence driving in both galaxies is primarily solenoidal on average for the regions where we find valid results. It is important to emphasise that these are lower limits of $b$, as we assume negligible magnetic fields due to the absence of magnetic-field measurements in the galaxies. We discuss the implications of this assumption (and others) later in Sec.~\ref{sec:caveats}. However, there is significant spatial variation of $b$ in both galaxies, which is primarily driven by the variations in $\mach$. While the 16th-to-84th-percentile range of $b$ is somewhat greater in NGC7793 than in NGC1313 ($\pm0.24$ compared to $\pm 0.17$), the total range of $b$ is larger in NGC1313, showing that different regions within NGC1313 have significant variation. Indeed, both galaxies show local values of $b$ as high as $b\sim0.5-0.8$, in the highly compressive-driving regime.

\subsection{Spatial variation of turbulent parameters} \label{sec:SpatialVar}
The results presented in Figures~\ref{fig:NGC7793_NGC1313_b_field} and \ref{fig:NGC7793_NGC1313_full_field_PDFs} and Tab.~\ref{tab:RegionalVarDependencies} show that there are clear spatial fluctuations of the turbulent parameters $\sigmarho$, $\mach$, and $b$ across both galaxies. Here, we examine the local and radial variations in the different turbulent parameters.

\begin{figure*}
    \centering
    \includegraphics[width=\linewidth]{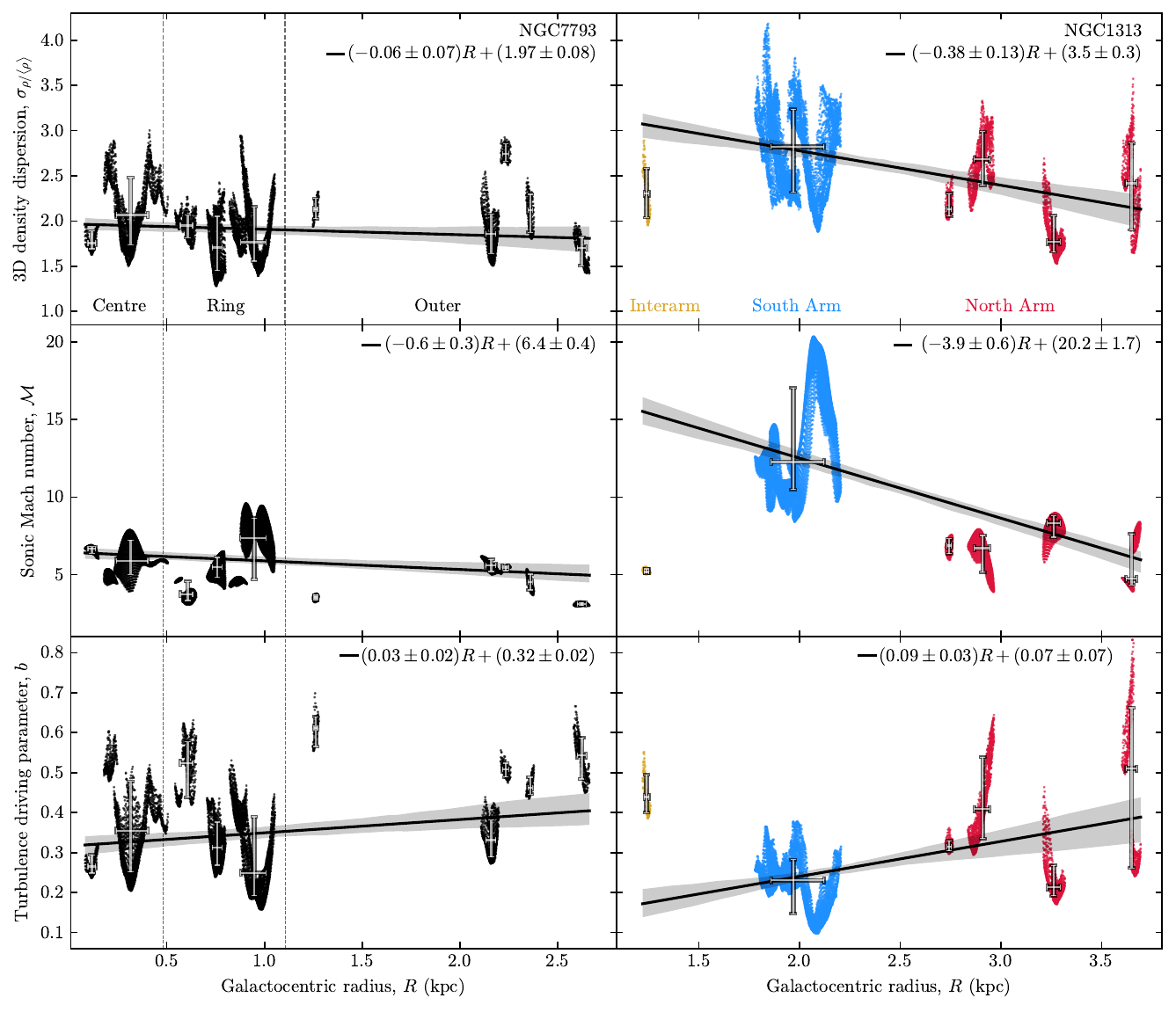}
    \caption{Galactocentric radial dependencies of $\sigmarho$ (top), $\mach$ (middle), and $b$ (bottom) in NGC7793 (left) and NGC1313 (right). The regions in NGC7793 (separated by vertical dashed lines) and NGC1313 (separated by different colours) are as defined in Sec.~\ref{sec:LocVar} and Fig.~\ref{fig:NGC7793_NGC1313_b_field}. The markers are binned radially by clusters of data, where the centre of the markers are the median values of the bin, and the 16th and 84th percentile values of the respective turbulent parameter and the radial distance shown by the tick markers for each clustered region of data. The fluctuations of the markers represents the significant spatial variability of the turbulent parameters. We perform a linear fit of all parameters and present the best-fit parameters together with their confidence interval in the top right corner of each panel. The relative percentage uncertainties of each measurement are listed in Tab.~\ref{tab:RegionalVarDependencies}. This figure shows both the global and local fluctuations in the turbulent parameters, highlighting the significant spatial variance of the turbulent parameters in different environments. We note specifically the variance in $b$, indicating various turbulent regimes coexisting in each galaxy.}
    \label{fig:RadDependencies}
\end{figure*}

\subsubsection{Local variations} \label{sec:LocVar}
Having examined the global spatial variations, we now explore the local variations of the turbulence by dividing each galaxy into different regions. The regions are shown in Fig.~\ref{fig:NGC7793_NGC1313_b_field} by the dashed contours. For NGC7793, the regions are separated by concentric annuli into Centre, Ring, and Outer, following the definitions in \citet{Finn_2024_b}. For NGC1313, the regions correspond to distinct morphological features: the North Arm, South Arm, Interarm, and Bar.

The measured turbulent parameters of these regions are listed in Tab.~\ref{tab:RegionalVarDependencies}. In NGC7793, we find that the Ring has the most solenoidally-driven turbulence ($b\geq0.29(\pm0.07)^{+0.11}_{-0.08}$), and the Outer region has the most compressively-driven turbulence ($b\geq0.41(\pm0.11)^{+0.12}_{-0.11}$) with the Centre having a turbulence driving more similar to the Outer region ($b\geq0.37(\pm0.10)^{+0.14}_{-0.12}$). We note that the median value of $b$ in the Outer and Centre regions are still in the mixed-to-compressive regime of turbulence driving. In NGC1313, the South Arm shows more solenoidal driving ($b\geq0.23(\pm0.05)^{+0.05}_{-0.08}$), and the Interarm region mostly mixed-to-compressive driving ($b\geq0.44(\pm0.14)^{+0.06}_{-0.04}$). While we do not find any results in the Bar region after the stringent data quality cuts, the North Arm shows a more solenoidal driving ($b\geq0.33(\pm0.08)^{+0.20}_{-0.12}$). 

There is also significant spatial variation within the local regions themselves, which is quantified by the 16th-to-84th percentile range of each value in Tab.~\ref{tab:RegionalVarDependencies}. The maximum 16th-to-84th-percentile spatial variation of $b$ is observed in the North Arm of NGC1313, extending from predominantly solenoidal driving ($b\lesssim 0.2$) to predominantly compressive driving ($b\gtrsim 0.5$).

The results in Tab.~\ref{tab:RegionalVarDependencies} suggest that the variations in $b$ are primarily driven by the variations in $\mach$, especially in the case of NGC1313. For example, the South Arm of NGC1313 shows $\sigmarho=2.8(\pm0.2)^{+0.4}_{-0.5}$, the highest of the galaxy. Conversely, the Interarm region has the value ($\sigmarho=2.3(\pm0.3)^{+0.3}_{-0.3}$), meaning one could expect a higher value of $b$ in the South Arm, and a lower value in the Interarm (by Equation~\ref{eqn:b_def}). Contrarily, $b$ is highest in the Interarm, and lowest in the South Arm, as  $\mach$ is proportionally decreased/increased more than $\sigmarho$.

\subsubsection{Radial trends} \label{sec:RadTrends}
We have explored the local and global spatial variance of the turbulent parameters. We now investigate the radial trends in Fig.~\ref{fig:RadDependencies}, which shows the galactocentric radial variation of the three turbulence parameters: the 3D turbulent density dispersion ($\sigmarho$), the sonic Mach number ($\mach$), and the turbulence driving parameter ($b$). Overall, we observe weak but systematic radial trends: $b$ tends to increase with galactocentric radius, while both $\sigmarho$ and $\mach$ exhibit mild declines. These patterns are more pronounced in NGC1313 than in NGC7793, which we compare more thoroughly in Sec.~\ref{sec:CompResults}. 

To assess the significance of the radial trends, we perform a bootstrap analysis of the linear regressions. For each of 100~bootstrap trials, we take a random selection of $0.5\%$ of the original data points, and re-fit each of those sub-samples. At each radial bin we collect the distribution of fitted values over all trials. The shaded confidence interval in Fig.~\ref{fig:RadDependencies} reflects the 16th-to-84th-percentile range in each radial bin, providing an estimate of the one-sigma confidence interval. In NGC7793, all radial trends are weak in that only $\mach$ shows a statistically significant dependence on radius, with a slight radial decline. NGC1313 shows clearer radial trends: $\sigmarho$ and $\mach$ decrease with radius, while $b$ increases, on average. We measure $b\geq(0.09\pm0.03)R+(0.07\pm0.07)$, where $R$ is the galactocentric radius (in units of kpc).

The radial increase in $b$ aligns with a scenario in which solenoidal turbulence dominates in the dense, central regions of galaxies, which are subject to strong shearing motions, while compressive driving becomes more prominent in the lower-density outskirts. This pattern is consistent with findings in the Milky Way, where clouds in the Central Molecular Zone (CMZ) exhibit more solenoidal driving compared to those in the Galactic disc \citep{Federrath_2016_Brick,Brunt_2010_taurus,Ginsburg_2013,Kainulainen_2017,Menon_2020}.

All radial fits in Fig.~\ref{fig:RadDependencies} exhibit substantial scatter, reflecting significant local variation in the turbulence parameters. Although NGC1313 shows weak radial trends, the broad dispersion around the fitted relations suggests that local environmental conditions exert a stronger influence. Our results therefore suggest that turbulence properties, particularly the driving parameter $b$, are primarily governed by local processes and should be measured and interpreted accordingly.

\subsection{Comparison of NGC7793 and NGC1313} \label{sec:CompResults}

The turbulent parameters differ for NGC1313 and NGC7793, likely reflecting the underlying morphological distinctions. NGC7793 is classified as a flocculent spiral with minimal coherent structure, whereas NC1313 exhibits well-defined features, including prominent spiral arms (North and South Arms) and a central bar. 

As discussed in Sec.~\ref{sec:RadTrends}, the radial trends in the turbulence parameters are more pronounced in NGC1313 than in NGC7793. This is likely due to the more well-defined morphological features in NGC1313, representing significantly different environments even on large scales. Accordingly, stronger radial trends emerge as the systematic variations in environments induce systematic variance in the turbulent conditions. In contrast, NGC7793 lacks such clear large-scale morphological differences, and therefore exhibits less radial variations in the turbulent parameters.

We also note that NGC1313 exhibits a lower global median value of $b$, indicating more solenoidally-driven turbulence on average compared to NGC7793 (or that NGC1313 is more magnetised). However, this global value is heavily dominated by the South Arm region, which accounts for over $70\%$ of the valid measurements in NGC1313. As a result, all global statistics of NGC1313, including their spatial variations, are dominated by this single region. In contrast, a larger global spatial variation of $b$ is observed in NGC7793, as measured by the 16th-to-84th-percentile variation listed in Tab.~\ref{tab:RegionalVarDependencies}. However, the North Arm region of NGC1313 shows the largest spatial variation of $b$ in all the regions, and when comparing both galaxies. These findings highlight that while NGC7793 shows greater global variation in $b$, NGC1313 contains regions with substantial internal scatter that are not reflected in the global metrics. This underscores the need to interpret galaxy-wide values with caution when regional sampling is uneven.

\section{Discussion}
In this section, we examine the implications of our findings, emphasising key features and their relevance to future investigations into turbulence and star formation in galaxies. We also outline the caveats and assumptions in the analysis, and outline efforts taken to minimise the impact of these assumptions. 

\subsection{Relevance for star formation}
As discussed in Sec.~\ref{sec:NGC1313_intro}, NGC1313 is undergoing a satellite interaction in the south-west. As shown in Fig.~\ref{fig:NGC7793_NGC1313_b_field}, most valid measurements in the South Arm region lie near this region, coinciding with the known starburst site associated with the interaction. Previous studies have shown that the encounter has disrupted the HI velocity field and likely also the CNM and molecular gas \citep{Sandage_1979,Peters_1994,Silva-Villa_2012,Hernandez_2022}. These disturbances create more chaotic conditions, which may explain the elevated values of $\sigmarho$ and $\mach$, along with the pronounced spatial variation in $\mach$ relative to other regions. It is important to acknowledge the statistical uncertainty in the individual kernel measurements as listed in Tab.~\ref{tab:RegionalVarDependencies}, but with the repeated kernel measurements within regions (see Sec.~\ref{sec:uncertainty_prop}) there is an increase in the statistical significance of the spatial variation of the parameters.

In this environment, the increase in $\mach$ dominates, leading to the lowest measured value of $b$ across both galaxies. However, at the same time, the SFR is enhanced in this region, likely due to the interaction. This is somewhat surprising, considering that previous studies have shown that, when all other parameters are identical (including the total gas mass, density, etc.), more compressive turbulence driving (i.e., higher $b$) produces higher SFRs (by about an order of magnitude) compared to solenoidal driving \citep{Federrath_2012}. However, $b$ is not the sole factor governing the SFR. \citet{Federrath_2012} identified four key parameters that regulate the $\SFRff$: the virial parameter ($\avir$), the Mach number ($\mach$), the turbulence driving parameter ($b$), and the ratio of thermal to magnetic pressure, plasma $\beta$. Thus, while $b$ may be relatively low, a low virial parameter or a large Mach number can lead to high SFR\footnote{While we do not currently have information about plasma $\beta$ in these galaxies, it is worth noting that the magnetic field plays another crucial role in regulating the SFR \citep{PadoanNordlund2011,Federrath_2015}.}.

The total SFR also depends on the total available gas mass and the absolute density. The merger may have brought extra gas in and/or led to an overall increase in density in the South Arm region, both of which would increase the total SFR -- not necessarily, however, the $\SFRff$, which is the fraction of available gas collapsing within a freefall time ($\tff$), and is primarily governed by $\avir$, $\mach$, $b$, and $\beta$. Indeed, neither $\tff$ nor the total gas mass is uniform across NGC1313. Therefore, regions with shorter local freefall times can exhibit elevated SFRs even when $\SFRff$ or $b$ are relatively lower.

In summary, we find significant spatial variation of the turbulent parameters across the galaxies. Consequently, it is not appropriate to assume a uniform turbulence driving mode across the face of a galaxy -- particularly given the substantial scatter in $b$ across different regimes. This is especially critical for the turbulence driving parameter $b$, as many previous studies of star formation have relied on the assumption of a uniform driving mode across diverse environments \citep[e.g.,][]{Hennebelle_2011, Krumholz_2012, Federrath2013sflaw, SalimFederrathKewley2015}. This assumption has largely been necessitated by the lack of robust, spatially resolved measurements of $b$, which is the purpose of this study.

\subsection{Caveats} \label{sec:caveats}
Our analysis incorporates several assumptions and data quality thresholds that may influence the quantitative values of the turbulent parameters. We argue that these do not affect the qualitative conclusions of this study, specifically the spatial variation of the turbulence driving parameter and other turbulent parameters. Below, we outline the key caveats and the steps taken to mitigate their impact.

\subsubsection{Uniform sound speed} \label{sec:soundspeedcav}
We adopt a fixed isothermal sound speed of $\cs = 0.6 \pm 0.3\,\mathrm{km\,s^{-1}}$, corresponding to gas temperatures in the range $10$--$150\,$K, consistent with an evenly mixed ISM comprising both CNM and molecular gas \citep[e.g.,][]{Ao_2013}. Since
$b\propto\mach^{-1}\propto\cs\propto\sqrt{T},$
spatial temperature variations could, in principle, induce systematic shifts in $b$. As discussed in Sec.~\ref{sec:soundspeed}, our measurements directly probe a mixture of CNM and molecular gas regions, so we do not expect significant temperature fluctuations when sampling similar environments. Nevertheless, we adopt a conservative uncertainty of $50\%$ for the sound speed, reflecting the approximations inherent in assuming a constant $\cs$, and this uncertainty is propagated through all subsequent results.

\subsubsection{S/N and kernel selection}
We adopt a conservative S/N threshold of 5 to ensure robust turbulent measurements (see Sec.~\ref{sec:S/NThresh}). \citet{Gerrard_2023} have explored the effects of different S/N threshold selections in their Appendix.~A, and figures~A1 and~A2. While higher thresholds improve accuracy, they significantly reduce the number of valid kernels available for analysis (see Sec.~\ref{sec:TurbKernels}). Consequently, our full-field maps (Fig.~\ref{fig:NGC7793_NGC1313_b_field}) yield relatively sparse coverage, limiting our ability to probe more environments within each galaxy. Nevertheless, despite the reduced spatial sampling, the uncertainties associated with each measurement (Tab.~\ref{tab:RegionalVarDependencies}) are fully propagated into our error budgets.

\subsubsection{Magnetic fields} \label{sec:magfieldcav}
We do not account for magnetic pressure or magnetically induced anisotropy in our measurement of $b$. In magnetised turbulence, the effective driving parameter would shift towards more compressive values \citep[e.g.,][]{Federrath_2016_Brick} if the magnetic field is taken into account. As such, our measured values of $b$ should be interpreted as lower limits. A reasonable range of values to be expected for $\beta$ is $0.1\lesssim \beta \lesssim 1$ under typical ISM conditions \citep[e.g.,][]{FalgaroneEtAl2008, Inoue_2012, Rainer_2015, Ferriere_2020}. This range corresponds to an upward scaling of $b$ by factors of $1.4$ to at most $3.3$ (see Eq.~\ref{eqn:b_def_magnetic}). We note that $\beta\sim0.1$ represents rather extreme conditions with very strong magnetic fields. In the absence of direct measurements of $\beta$, all analyses in this paper assume $\beta\to\infty$, and therefore the reported $b$ values should be considered lower limits.

\subsubsection{Non-turbulent influences on the velocity dispersion}
Our use of the second moment, $M_2$, to estimate $\sigma_{v,1\D}$ does not account for large-scale systematic motions (e.g., rotation or shear), which can artificially inflate $\mach$ and thereby reduce the inferred value of $b$.

\subsubsection{Isotropy of the density power spectrum} \label{sec:IsoDensPowerField}
The \citet{Brunt_2010} method (Sec.~\ref{sec:vol_disps}) assumes that the density power spectra are isotropic in all directions. Given that we consider local kernel scales, this is a reasonable approximation. Random inspections verify this assumption, including the example shown in Fig.~\ref{fig:PowerSpectra} in Appendix~\ref{sec:PowerSpectEG}, which corresponds to the kernel from Fig.~\ref{fig:Mom0_kernel_subtr}. We fit ellipses to different intensity level contours of the $2\D$ power spectra in Fig.~\ref{fig:Mom0_kernel_subtr}, which yields a maximum axis ratio (largest to the smallest radius) of $\lesssim1.2$, consistent with nearly isotropic density power spectra \citep{Federrath_2016_Brick}. While \citet{Brunt_2010} report a $10\%$ uncertainty for a perfectly isotropic field, we adopt a conservative $20\%$ uncertainty to account for the slight anisotropy present in our fields. This uncertainty is incorporated and propagated through all results (Sec.\ref{sec:uncertainty_prop}; Tab.~\ref{tab:uncertaintyProp}).

\subsubsection{Universality of $b$}
Some measurements yield $b<1/3$, i.e., below the theoretical limit of purely solenoidal driving. This can be attributed to a combination of observational and theoretical factors. Neglecting magnetic fields is likely the primary cause of these low $b$ values (see Sec.~\ref{sec:magfieldcav}). Additionally, the turbulence isolation process (Sec.~\ref{sec:LowPassFilter}) may not fully remove non-turbulent contributions from the density and/or velocity fields. Nevertheless, inaccuracies due to imperfect turbulence isolation are expected to be minor, as the resulting column density PDFs exhibit typical features of fully-developed turbulent flows (see Fig.~\ref{fig:PDFs}).

\subsubsection{Summary}
The caveats discussed above outline the main systematic uncertainties in our study. Despite these limitations, the spatial variations and relative differences of $b$ -- both between regions and between galaxies -- remain significant, providing a solid foundation for future high-sensitivity observations and MHD extensions of this work. Given the caveats discussed and the large relative uncertainties presented in Tab.~\ref{tab:RegionalVarDependencies}, we emphasise the importance of spatial variability in $b$ and other turbulent parameters, rather than their absolute calibration. The uncertainty in the measurement of $b$ exceeds $\sim70\%$ for the kernel measurements of both galaxies, precluding definitive conclusions about individual kernel absolute measurements. Nonetheless, the regional spatial variation of the turbulent parameters confirm the significant spatial variation of the turbulent parameters. The observed spatial variation in $b$ demonstrates that turbulence driving is not uniform across different galactic environments. 

\subsection{Future Direction}
The methodology developed in this work opens several avenues to advance our understanding of turbulence driving in galaxies. Applying this framework to larger galaxy samples, such as those in the PHANGS survey \citep{Leroy_2021_Pipeline, Leroy_2021_SFGals}, will allow statistical exploration of how turbulence driving varies with morphology, mass, and star formation activity. At higher redshift, spatially resolved IFU observations, such as SDSS-V LVM \citep{Drory_2024}, may help enable the first direct tests of whether turbulence driving evolves with cosmic time. In parallel, incorporating turbulence driving into chemo-dynamical models will clarify the role of turbulence in regulating star formation, feedback, and metal mixing. Ultimately, it is the combination of high-resolution observations with state-of-the-art theoretical predictions that will establish the role of turbulence driving as a key parameter linking the physics of the interstellar medium to the evolution of galaxies. In this future research we emphasise the importance of magnetic field measurements as well as high spatial resolution observations to investigate the local scales of turbulent variance.

\section{Conclusions} \label{sec:conclusion}
We present the first extragalactic, spatially resolved maps of the turbulence driving parameter $b$ in the nearby galaxies NGC7793 and NGC1313, using ALMA \CO observations and a roving-kernel analysis technique (Fig.~\ref{fig:NGC7793_NGC1313_b_field}). This work establishes an observational framework for linking small-scale turbulence physics to large-scale galactic environment and star formation regulation. Our results indicate four key conclusions: 
\begin{enumerate}
    \item 
    The lower limits of the turbulence driving parameter, $b$, are in the solenoidal driving regime, with values of $b\geq0.33(\pm0.05)^{+0.14}_{-0.10}$ in NGC7793 and $b\geq0.24(\pm0.03)^{+0.10}_{-0.07}$ in NGC1313. These results are consistent with the idea that large-scale shear and rotation in galactic discs act as primary sources of turbulence driving, favouring solenoidal forcing. We note that these are strictly lower limits of $b$, as they do not include the influence of magnetic fields (cf.~Sec.~\ref{sec:magfieldcav}). 
    \item 
    Despite these global averages, both systems exhibit substantial spatial variation. We identify localised regions where turbulence driving is strongly compressive ($b\gtrsim0.5$), demonstrating that the turbulence mode is not uniform within a galaxy. These compressive regions often occur in association with enhanced density dispersions or disturbed morphologies, suggesting that localised feedback, cloud–cloud collisions, or tidal interactions can temporarily shift the turbulence regime away from solenoidal driving. 
    \item 
    NGC1313, which shows evidence of a recent interaction, displays higher turbulent Mach numbers and stronger density fluctuations than the more quiescent NGC7793. NGC1313 also shows a clear radial gradient, with solenoidal modes dominating in the centre and more compressive modes emerging in the outer disc. This suggests that central regions are regulated by shear and differential rotation, while the outskirts are more sensitive to accretion, tidal effects, and large-scale flows. In contrast, NGC7793 shows a more regular, nearly flat radial profile, consistent with its undisturbed morphology.
    \item These findings highlight that turbulence driving is not a single global property but varies systematically across different galactic environments, and as such, the turbulent parameters $\sigmarho$, $\mach$ and $b$ should be treated as locally-defined variables. This variation has direct consequences for theories of star formation regulation: while compressive driving can promote rapid local collapse, solenoidal driving suppresses it, and the balance between the two must be considered alongside Mach number, virial parameter, and the plasma-$\beta$ term. Our results provide a new observational handle on this balance, showing that galactic environment sets the stage for which driving mode dominates in different regions.
\end{enumerate}

In summary, turbulence in galaxies cannot be captured by a single driving mode. Our results indicate that the driving mode of turbulence -- alongside other key turbulent parameters -- cannot be assumed constant across a galaxy. They vary across space, scale, and environment, with various turbulence driving emerging in specific galactic conditions. The significant regional variance we observe must be accounted for in any future models of star formation, molecular cloud evolution, and our understanding of gas kinematics of the ISM. By introducing a methodology to map turbulence driving in external galaxies and the first measurements of the spatial variation of the turbulence driving, this work opens a new observational window onto the physics of the ISM and provides a foundation for connecting local turbulence processes to the evolution of galaxies as a whole.

\section*{Acknowledgements}
We thank the anonymous referee for their valuable comments, which helped improve this work. We further thank M.~Finn for providing the foundational analysis and moment maps of NGC7793 and NGC1313, and for advice on the data processing of the galaxies. We also thank Janet Tang for helpful discussions on the analysis methods. K.G.~is supported by the Australian Research Council through the Discovery Early Career Researcher Award (DECRA) Fellowship (project number~DE220100766) funded by the Australian Government. C.F.~acknowledges funding provided by the Australian Research Council (Discovery Project grants~DP230102280 and~DP250101526), and the Australia-Germany Joint Research Cooperation Scheme (UA-DAAD). We further acknowledge high-performance computing resources provided by the Leibniz Rechenzentrum and the Gauss Centre for Supercomputing (grants~pr32lo, pr48pi and GCS Large-scale project~10391), the Australian National Computational Infrastructure (grant~ek9) and the Pawsey Supercomputing Centre (project~pawsey0810) in the framework of the National Computational Merit Allocation Scheme and the ANU Merit Allocation Scheme. This paper makes use of ALMA data ADS/JAO.ALMA\#2015.1.00782.S. ALMA is a partnership of ESO (representing its member states), NSF (USA), and NINS (Japan), together with NRC (Canada), NSC and ASIAA (Taiwan), and KASI (Republic of Korea), in cooperation with the Republic of Chile. The Joint ALMA Observatory is operated by ESO, AUI/NRAO, and NAOJ. The National Radio Astronomy Observatory is a facility of the National Science Foundation operated under cooperative agreement by Associated Universities, Inc.

\section*{Data Availability}
The data underlying this article are available from the ALMA archive, found at \url{https://almascience.nrao.edu/aq/}.



\bibliographystyle{mnras}
\bibliography{references,federrath}



\newpage
\appendix

\section{Power spectra example} \label{sec:PowerSpectEG}
\begin{figure} 
    \includegraphics[width=\linewidth]{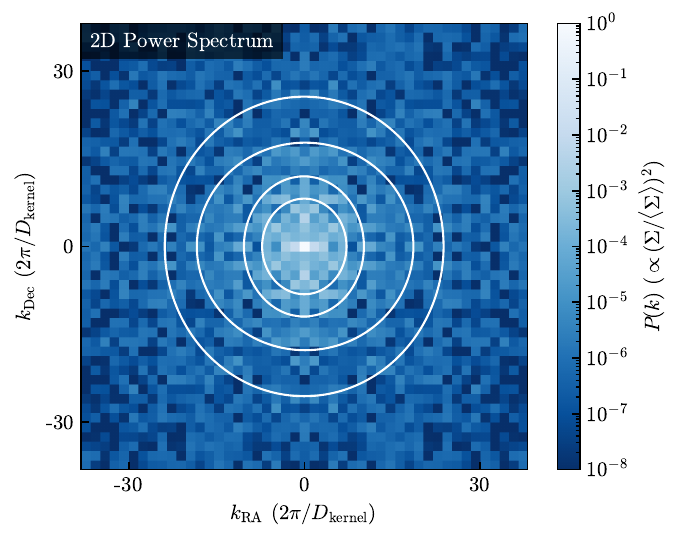}
    \caption{
    Column density power spectrum of the same example kernel as in Fig.~\ref{fig:Mom0_kernel_subtr}. We fit ellipses to the contours corresponding to $10^{-4}$, $10^{-5}$, $10^{-6}$ and $10^{-7}$, demonstrating a reasonable level of isotropy. The ratios of the major-to-minor axes of the ellipses for the different contours are 1.13, 1.17, 1.04 and 1.07, from high to low power respectively.}
    \label{fig:PowerSpectra}
\end{figure}

Fig.~\ref{fig:PowerSpectra} shows the power spectrum of the column density field for the same example kernel as in Fig.~\ref{fig:Mom0_kernel_subtr}, demonstrating a reasonable level of isotropy, which is reflected in the uncertainty estimate for $\mathcal{R}^{1/2}$ Brunt factor in Tab.~\ref{tab:uncertaintyProp} (see also Sec.~\ref{sec:IsoDensPowerField}). The maximum axis ratio of the fitted ellipses is $\lesssim1.2$, making the fields nearly isotropic and hence the uncertainty in the $\mathcal{R}^{1/2}$ Brunt factor is $\lesssim20\%$, as discussed in Sec.~\ref{sec:IsoDensPowerField}. 

\section{S/N threshold and the brunt factor} \label{sec:bruntR_interpimpacts}
\begin{figure*}
    \centering
    \includegraphics[width=0.99\linewidth]{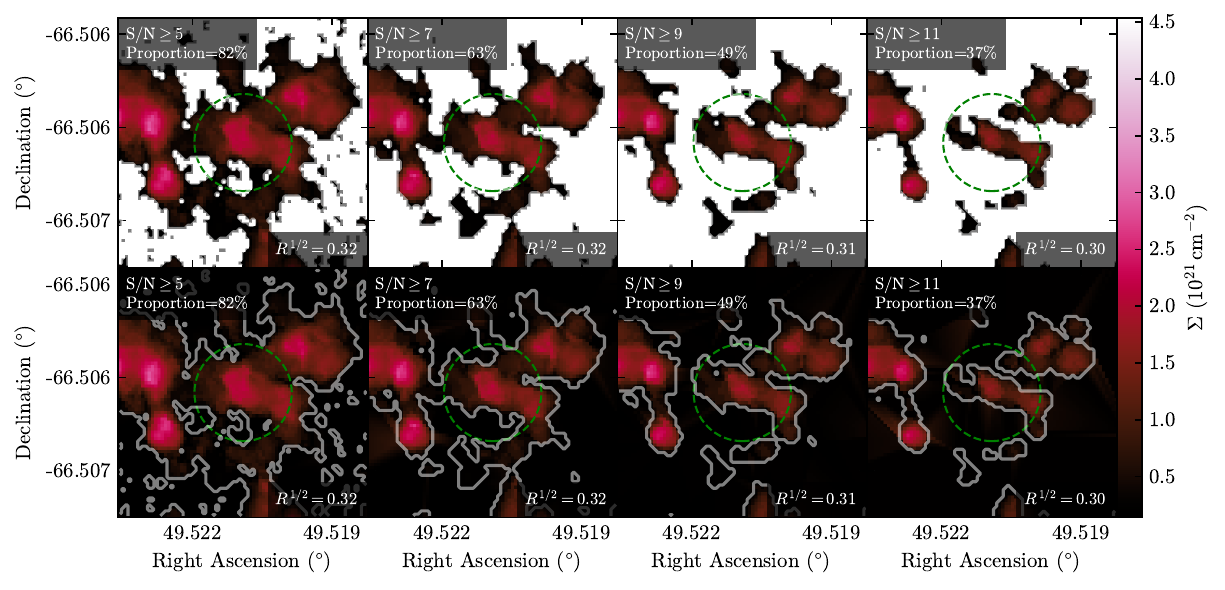}
    \caption{The influence of the interpolation process on the $R^{1/2}$ Brunt factor. The top row shows the data without interpolation and white pixels are where there are no velocity channels reaching the designated S/N. The bottom row shows the same, but with the interpolated regions shown. From left to right, the S/N threshold is increasing, and the associated kernel-weighted proportion of pixels is decreasing, as labelled. We find that the $R^{1/2}$ Brunt factor varies only minimally ($\leq4\%$), even when $\sim50\%$ of the spaxels are interpolated.}
    \label{fig:InterpBruntREffects}
\end{figure*}

Here, we discuss the influence of the interpolation process detailed in Sec.~\ref{sec:interp}, specifically with regards to the uncertainties induced in the $R^{1/2}$ Brunt factor. In particular, we reference Fig.~\ref{fig:InterpBruntREffects} where we isolate a region of high S/N in NGC1313, and amplify the S/N threshold on the data, an approximation for reducing data quality and increasing the amount of interpolated data. We find that the change in $R^{1/2}$ is less than $4\%$ even when down to a kernel-weighted $\sim50\%$ of the spaxels being filled (our required minimum), which is well incorporated into the $20\%$ uncertainty assigned to $R^{1/2}$.

\bsp	
\label{lastpage}
\end{document}